\documentclass[journal]{IEEEtran}

\usepackage[numbers,sort&compress]{natbib}
\usepackage{url}
\usepackage{bm}
\usepackage{amsfonts}
\usepackage{amsmath}
\usepackage{amssymb}
\usepackage{amsthm}
\usepackage{color}
\usepackage{enumerate}

\usepackage{array,color}
\usepackage{tabularx}
\usepackage{multirow}
\usepackage{booktabs}
\usepackage{makecell}  

\usepackage{graphicx}  
\usepackage{subfig}
\usepackage{epstopdf}
\usepackage{graphics}
\usepackage{epsfig}
\usepackage{psfrag}
\usepackage{overpic}

 \usepackage{stfloats}
 \usepackage{float}   

 \usepackage{indentfirst} 
\usepackage{gensymb}

\begin{document}

\newcommand {\Data} [1]{\mbox{${#1}$}}  

\newcommand {\DataN} [2]{\Data{\Power{{#1}}{{{#2}}}}}  
\newcommand {\DataIJ} [3]{\Data{\Power{#1}{{{#2}\!\times{}\!{#3}}}}}  

\newcommand {\DatassI} [2]{\!\Data{\Index{#1}{\!\Data 1},\!\Index{#1}{\!\Data 2},\!\cdots,\!\Index{#1}{\!{#2}}}}  
\newcommand {\DatasI} [2]{\Data{\Index{#1}{\Data 1},\Index{#1}{\Data 2},\cdots,\Index{#1}{#2},\cdots}}   
\newcommand {\DatasII} [3]{\Data{\Index{#1}{{\Index{#2}{\Data 1}}},\Index{#1}{{\Index{#2}{\Data 2}}},\cdots,\Index{#1}{{\Index{#2}{#3}}},\cdots}}  

\newcommand {\DatasNTt}[3]{\Data{\Index{#1}{{#2}{\Data 1}},\Index{#1}{{#2}{\Data 2}},\cdots,\Index{#1}{{#2}{#3}}} } 
\newcommand {\DatasNTn}[3]{\Data{\Index{#1}{{\Data 1}{#3}},\Index{#1}{{\Data 2}{#3}},\cdots,\Index{#1}{{#2}{#3}}} } 

\newcommand {\Vector} [1]{\Data {\mathbf {#1}}}
\newcommand {\Rdata} [1]{\Data {\hat {#1}}}
\newcommand {\Tdata} [1]{\Data {\tilde {#1}}} 
\newcommand {\Udata} [1]{\Data {\overline {#1}}} 
\newcommand {\Fdata} [1]{\Data {\mathbb {#1}}} 
\newcommand {\Prod} [2]{\Data {\prod_{\SI {#1}}^{\SI {#2}}}}  
\newcommand {\Sum} [2]{\Data {\sum_{\SI {#1}}^{\SI {#2}}}}   
\newcommand {\Belong} [2]{\Data{ {#1} \in{}{#2}}}  

\newcommand {\Abs} [1]{\Data{ \lvert {#1} \rvert}}  
\newcommand {\Mul} [2]{\Data{ {#1} \times {#2}}}  
\newcommand {\Muls} [2]{\Data{ {#1} \! \times \!{#2}}}  
\newcommand {\Mulsd} [2]{\Data{ {#1} \! \cdot \!{#2}}}  
\newcommand {\Div} [2]{\Data{ \frac{#1}{#2}}}  
\newcommand {\Trend} [2]{\Data{ {#1}\rightarrow{#2}}}  
\newcommand {\Sqrt} [1]{\Data {\sqrt {#1}}} 
\newcommand {\Sqrnt} [2]{\Data {\sqrt[2]{#1}}} 

\newcommand {\Power} [2]{\Data{ {#1}^{\TI {#2}}}}  
\newcommand {\Index} [2]{\Data{ {#1}_{\TI {#2}}}}  

\newcommand {\Equ} [2]{\Data{ {#1} = {#2}}}  
\newcommand {\Equs} [2]{\Data{ {#1}\! =\! {#2}}}  
\newcommand {\Equss} [3]{\Equs {#1}{\Equs {#2}{#3}}}  

\newcommand {\Equu} [2]{\Data{ {#1} \equiv {#2}}}  

\newcommand {\LE}[0] {\leqslant}
\newcommand {\GE}[0] {\geqslant}
\newcommand {\NE}[0] {\neq}
\newcommand {\INF}[0] {\infty}
\newcommand {\MIN}[0] {\min}
\newcommand {\MAX}[0] {\max}

\newcommand {\Funcfx} [2]{\Data{ {#1}({#2})}}  
\newcommand {\Funcfzx} [3]{\Data{ {\Index {#1}{#2}}({#3})}}  
\newcommand {\Funcfnzx} [4]{\Data{ {\Index {\Power{#1}{#2}}{#3}}({#4})}}  
\newcommand {\SI}[1] {\small{#1}}
\newcommand {\TI}[1] {\tiny {#1}}
\newcommand {\Text}[1] {\text {#1}}

\newcommand {\VtS}[0]{\Index {t}{\Text {s}}}
\newcommand {\Vti}[0]{\Index {t}{i}}
\newcommand {\Vt}[0]{\Data {t}}
\newcommand {\VLES}[1]{\Index {\tau} {\SI{\Index {}{ \Text{#1}}}}}
\newcommand {\VLESmin}[1]{\Index {\tau} {\SI{\Index {}{ \Text{min\_\Text{#1}}}}}}
\newcommand {\VAT}[0]{\Index {\Vector A}{\Text{Time}}}
\newcommand {\VPbus}[1]{\Index {P}{\Text{Node-}{#1}}}
\newcommand {\VPbusmax}[1]{\Index {P}{\Text{max\_Node-}{#1}}}

\newcommand {\EtS}[2]{\Equs {\Index {t}{\Text {s}}}{#1} {#2}}
\newcommand {\Eti}[2]{\Equs {\Index {t}{i}}{#1} {#2}}
\newcommand {\Et}[2]{\Equs {t}{#1} {#2}}
\newcommand {\EMSR}[2]{\Equs {\Index {\tau} {\SI{\Index {}{ \Text{MSR}}}}}{#1} {#2}}
\newcommand {\EMSRmin}[2]{\Equs {\Index {\tau} {\SI{\Index {}{ \Text{MSR}}}}}{#1} {#2}}

\newcommand {\EAT}[2]{\Equs {\Index {\Vector A}{\Text{Time}}}{#1} {#2}}
\newcommand {\EPbus}[3]{\Equs {\Index {P}{\Text{Node-}{#1}}}{#2} \Text{ #3}}
\newcommand {\EPbusmax}[3]{\Equs {\Index {P}{\Text{max\_Node-}{#1}}}{#2} \Text{ #3}}

\newcommand {\Vgam}[1]{\Index {\gamma}{#1}}
\newcommand {\Egam}[2]{\Equs {\Vgam{#1}}{#2}}

\newcommand {\Emu}[2]{\Equs {{\mu}{#1}}{#2}}
\newcommand {\Esigg}[2]{\Equs {{\sigma}^2{#1}}{#2}}

\newcommand {\Vlambda}[1]{\Index {\lambda}{#1}}

\newcommand {\VV}[1]{\Index {\Vector V}{#1}}
\newcommand {\Vv}[1]{\Index {\Vector v}{#1}}
\newcommand {\Vsv}[1]{\Index {v}{#1}}
\newcommand {\VX}[1]{\Index {\Vector X}{#1}}
\newcommand {\VsX}[1]{\Index {X}{\SI{\Index {}{#1}}}}
\newcommand {\Vx}[1]{\Index {\Vector x}{\SI{\Index {}{#1}}}}
\newcommand {\Vsx}[1]{\Index {x}{\SI{\Index {}{#1}}}}
\newcommand {\VZ}[1]{\Index {\Vector Z}{#1}}
\newcommand {\Vz}[1]{\Index {\Vector z}{\SI{\Index {}{#1}}}}
\newcommand {\Vsz}[1]{\Index {z}{\SI{\Index {}{#1}}}}
\newcommand {\VIndex}[2]{\Index {\Vector {#1}}{#2}}
\newcommand {\VY}[1]{\Index {\Vector Y}{#1}}
\newcommand {\Vy}[1]{\Index {\Vector y}{#1}}
\newcommand {\Vsy}[1]{\Index {y}{\SI{\Index {}{#1}}}}

\newcommand {\VRV}[1]{\Index {\Rdata {\Vector V}}{#1}}
\newcommand {\VRsV}[1]{\Index {\Rdata {V}}{#1}}
\newcommand {\VRX}[1]{\Index {\Rdata {\Vector X}}{#1}}
\newcommand {\VRx}[1]{\Index {\Rdata {\Vector x}}{\SI{\Index {}{#1}}}}
\newcommand {\VRsx}[1]{\Index {\Rdata {x}}{\SI{\Index {}{#1}}}}
\newcommand {\VRZ}[1]{\Index {\Rdata {\Vector Z}}{#1}}
\newcommand {\VRz}[1]{\Index {\Rdata {\Vector z}}{\SI{\Index {}{#1}}}}
\newcommand {\VRsz}[1]{\Index {\Rdata {z}}{\SI{\Index {}{#1}}}}
\newcommand {\VTX}[1]{\Index {\Tdata {\Vector X}}{#1}}
\newcommand {\VTx}[1]{\Index {\Tdata {\Vector x}}{\SI{\Index {}{#1}}}}
\newcommand {\VTsx}[1]{\Index {\Tdata {x}}{\SI{\Index {}{#1}}}}
\newcommand {\VTsX}[1]{\Index {\Tdata {X}}{\SI{\Index {}{#1}}}}
\newcommand {\VTZ}[1]{\Index {\Tdata {\Vector Z}}{#1}}
\newcommand {\VTz}[1]{\Index {\Tdata {\Vector z}}{\SI{\Index {}{#1}}}}
\newcommand {\VTsz}[1]{\Index {\Tdata {z}}{\SI{\Index {}{#1}}}}
\newcommand {\VOG}[1]{\Vector{\Omega}{#1}}

\newcommand {\Sigg}[1]{\Data {{\sigma}^2({#1})}}
\newcommand {\Sig}[1]{\Data {{\sigma}({#1})}}

\newcommand {\Mu}[1]{\Data{{\mu} ({#1})}}
\newcommand {\Eig}[1]{\Data {\lambda}({\Vector {#1}}) }
\newcommand {\Her}[1]{\Power {#1}{\!H}}
\newcommand {\Tra}[1]{\Power {#1}{\!T}}

\newcommand {\VF}[3] {\DataIJ {\Fdata {#1}}{#2}{#3}}
\newcommand {\VRr}[2] {\DataN {\Fdata {#1}}{#2}}

\newcommand {\Tcol}[2] {\multicolumn{1}{#1}{#2} }
\newcommand {\Tcols}[3] {\multicolumn{#1}{#2}{#3} }
\newcommand {\Cur}[2] {\mbox {\Data {#1}-\Data {#2}}}

\newcommand {\VDelta}[1] {\Data {\Delta\!{#1}}}

\newcommand {\STE}[1] {\Fdata {E}{\Data{({#1})}}}
\newcommand {\STD}[1] {\Fdata {D}{\Data{({#1})}}}

\newcommand {\TestF}[1] {\Data {\varphi(#1)}}
\newcommand {\ROMAN}[1] {\uppercase\expandafter{\romannumeral#1}}

\def \FuncC #1#2{
\begin{equation}
{#2}
\label {#1}
\end{equation}
}

\def \FuncCC #1#2#3#4#5#6{
\begin{equation}
#2=
\begin{cases}
    #3 & #4 \\
    #5 & #6
\end{cases}
\label{#1}
\end{equation}
}

\def \Figff #1#2#3#4#5#6#7{   
\begin{figure}[#7]
\centering
\subfloat[#2]{
\label{#1a}
\includegraphics[width=0.23\textwidth]{#4}
}
\subfloat[#3]{
\label{#1b}
\includegraphics[width=0.23\textwidth]{#5}
}
\caption{\small #6}
\label{#1}
\end{figure}
}

\def \Figffb #1#2#3#4#5#6#7#8#9{   
\begin{figure}[#9]
\centering
\subfloat[#2]{
\label{#1a}
\includegraphics[width=0.23\textwidth]{#5}
}
\subfloat[#3]{
\label{#1b}
\includegraphics[width=0.23\textwidth]{#6}
}

\subfloat[{#4}]{
\label{#1c}
\includegraphics[width=0.48\textwidth]{#7}
}
\caption{\small #8}
\label{#1}
\end{figure}
}

\def \Figffp #1#2#3#4#5#6#7{   
\begin{figure*}[#7]
\centering
\subfloat[#2]{
\label{#1a}
\begin{minipage}[t]{0.24\textwidth}
\centering
\includegraphics[width=1\textwidth]{#4}
\end{minipage}
}
\subfloat[#3]{
\label{#1b}
\begin{minipage}[t]{0.24\textwidth}
\centering
\includegraphics[width=1\textwidth]{#5}
\end{minipage}
}
\caption{\small #6}
\label{#1}
\end{figure*}
}

\def \Figf #1#2#3#4{   
\begin{figure}[#4]
\centering
\includegraphics[width=0.48\textwidth]{#2}

\caption{\small #3}
\label{#1}
\end{figure}
}

\definecolor{Orange}{RGB}{249,106,027}
\definecolor{sOrange}{RGB}{251,166,118}
\definecolor{ssOrange}{RGB}{254,213,190}

\definecolor{Blue}{RGB}{008,161,217}
\definecolor{sBlue}{RGB}{090,206,249}
\definecolor{ssBlue}{RGB}{200,239,253}

\title{A Novel Data-Driven Situation Awareness Approach for Future Grids---Using Large Random Matrices for Big Data Modeling}

\author{Xing~He,  Lei~Chu,  Robert~C. Qiu,~\IEEEmembership{Fellow,~IEEE},  Qian~Ai,~\IEEEmembership{Senior Member,~IEEE},   Zenan~Ling
\thanks{This work was partly supported by National Natural Science Foundation of China (No. 61571296). \textcircled{c} 20xx IEEE. Personal use of this material is permitted. Permission  from IEEE must be obtained for all other uses, in any current or future  media, including reprinting/republishing this material for advertising or  promotional purposes, creating new collective works, for resale or  redistribution to servers or lists, or reuse of any copyrighted  component of this work in other works.}
}

\maketitle


\begin{abstract}
Data-driven approaches, when tasked with situation awareness,  are suitable for complex grids with \textit{massive datasets}. It is a challenge, however, to  efficiently turn these massive datasets into useful big data analytics. To address such a challenge, this paper, based on random matrix theory (RMT), proposes a data-driven approach. The approach models massive datasets as large random matrices; it is model-free and requiring no knowledge about physical model parameters. In particular, the large data dimension $N$ and the large time span $T$, from the spatial aspect and the temporal aspect respectively, lead to favorable results. The beautiful thing lies in that these linear eigenvalue statistics (LESs) built from data matrices follow Gaussian distributions for very general conditions, due to the \textit{latest breakthroughs} in probability on the central limit theorems of those LESs. Numerous case studies, with both simulated data and field data, are given to validate the proposed new algorithms.

\end{abstract}

\begin{IEEEkeywords}
Big data analytics, situation awareness, random matrix theory, linear eigenvalue statistics, statistical indicator
\end{IEEEkeywords}

\IEEEpeerreviewmaketitle

\section{Introduction}
\IEEEPARstart{S}{ituation} awareness (SA) is of great significance for power system operation, and a reconsideration of SA is essential for future grids~\cite{bda2016tsg}. These future grids are always huge in size and  complex in topology. Operating under a novel regulation, their management mode is much different~\cite{he2015arch}.   Data are more and more easily accessible, on the other hand, and data-driven approaches become natural for future grids.
Towards this vision, we are facing the following challenges:
\begin{itemize}
\item There are massive data in power grids. The so-called curse of dimensionality~\cite{moulin2004support} occurs inevitably.
\item The resource cost (time, hardware, human, etc.) for  extracting big data analytics should be tolerable.
\item For a massive data source, there often exist realistic ``bad'' data, e.g. the incomplete, the inaccurate, the asynchronous, and the unavailable. For system operations, decisions such as relay actions, should be highly reliable.
\end{itemize}

This paper is built upon our previous work in the last several years. See Section~\ref{sect:RelationshipPrevious} for details. Motivated for data mining, our line of research is based on the high-dimensional statistics. By high-dimensionality, we mean that the datasets are represented in terms of large random matrices. These data matrices can be viewed as data points in high-dimensional vector space---each vector is very long.

Data-driven approaches and data utilization for smart grids are  current stressing topics, as evidenced in the special issue of  ``Big Data Analytics for Grid Modernization''~\cite{bda2016tsg}. This special issue is most relevant to our paper in spirit. Several SA topics are discussed. We highlight the anomaly detection and classification~\cite{7460963, 7452675},  the estimation of active ingredients such as PV installations~\cite{7456317, 7347457}, and finally the online transient stability evaluation using real-time data~\cite{7445227}. 

In addition, we point out  research about the improvement in wide-area monitoring, protection and control (WAMPAC) and the utilization of PMU data~\cite {phadke2008wide,terzija2011wide,xie2012distributed}, together with the fault detection and location~\cite{jiang2012pmu, al2014fully}.
Xie et al. based on principal component analysis (PCA), propose an online application for early event detection by introducing a reduced dimensionality \cite {xie2014dimensionality}. The work has a special connection to our paper. Lim et al. study the quasi-steady-state operational problem relevant to the voltage instability phenomenon based on SVD using PMU data~\cite {lim2016svd}.

\subsection{Contributions of Our Paper}
Randomness is critical to future grids since rapid fluctuations in voltages and currents are ubiquitous. Often, these fluctuations exhibit Gaussian statistical properties~\cite {lim2016svd}. Our central interest in this paper is to model these rapid fluctuations using the framework of random matrix theory (RMT).  Our new algorithms are made possible due to the \textit{latest breakthroughs} in probability on the central limit theorems of the linear eigenvalue statistics (LESs)~\cite[Chapter 7]{qiu2015smart}. See~\cite{Qiu2016BigDataLRM} for a recent review.
\begin{enumerate}
\item  Starting from fundamental formulas of power systems, a theoretical justification is given for the validity of modeling complex grids as large random matrices. This data modeling framework ties together the RMT and the power system analysis. This part is basic in nature.
\item We study numerous basic problems including the technical route and applied framework, data-processing and relevant procedures, evaluation system and indicator sets, and the advantages over classical methodologies.
\item We make a comparison between RMT-based approach and PCA-based  one.
\item On the basis of  big data analytics, we study some power system applications: anomaly detection and location, empirical spectral density test, sensitivity analysis, statistical indicator system and its visualization, and, finally, robustness against asynchronous data.
\end{enumerate}

\subsection{Relationship to Our Previous Work}
\label{sect:RelationshipPrevious}
Our work~\cite{he2015arch}  is the first attempt to introduce the mathematical tool of RMT into power systems. Later, numerous papers demonstrate the power of this tool. Ring Law and Marchenko-Pastur (M-P) Law are regarded as the statistical foundation, and Mean Spectral Radius (MSR) is proposed as the high-dimensional indicator. Then we move forward to the second stage---paper \cite{he2015corr} studies the correlation analysis under the above framework. The concatenated matrix $\mathbf{A}_i$ is the object of interest. It consists of the basic matrix $\mathbf{B}$ and a factor matrix $\mathbf{C}_i$, i.e., $\mathbf{A}_i\!=\![\mathbf{B}; \mathbf{C}_i]$. In order to seek the sensitive factors, we compute the  advanced indicators that are based on  the LESs of these concatenated matrices $\mathbf{A}_i$. This study contributes to fault detection and location, line-loss reduction, and power-stealing prevention.
We also conduct analysis for power transmission equipment based on the same theoretical foundation \cite{yan2016transmission}.
Paper \cite{he2015les} is the third step in which the LES set is studied. Based on the LES set, a statistical and data-driven indicator system, rather than its deterministic and  model-based counterpart, is built to describe the system from a high-dimensional perspective. The robustness against spatial data error, precisely, data losses in the core area, is emphasized.

\subsection{Advantages of RMT-based Approach}
\label{AdvantagesofRMT}
The data-driven approach conducts analysis requiring no prior knowledge of the system topology, the unit operation/control mechanism, the causal relationship, etc.  Comparing with classical data-driven methodologies such as PCA-based method, the RMT-based counterpart has some unique advantages:\\
1) The massive dataset of power systems are in a high-dimensional vector space; the temporal variations ($T$ sampling instants) are simultaneously observed together with spatial variations ($N$ grid nodes). The extraction  of   information  from the above temporal-spatial variations is  a challenge that does not meet the prerequisites of most classical statistical algorithms.
Unifying time and space  through their ratio $c = T/N$, RMT deal with such kind of data mathematically rigorously. 
\\
2) The statistical indicator is generated from all the data in the form of matrix entries. This is not true to principal components---we really do not know the rank of the covariance matrix. The large size of the data enhances the robustness of the final decision against the bad data (inaccuracy, losses, or asynchronization), as well as those challenges in classical data-driven methods,  such as error accumulations and spurious correlations \cite{he2015corr}.\\
3) For the statistical indicator, a theoretical or empirical value can be obtained in advance. The statistical indicator such as LES follows a Gaussian distribution, and its variance is bounded \cite{shcherbina2011central} and decays very fast in the order of  $O({N^{-2}})$ for a given data dimension $N,$ say $N=118.$\\
4) We can flexibly handle heterogenous data to realize data fusion via matrix operations, such as the blocking \cite{he2015arch}, the sum \cite{zhang2015MassiveMIMO}, the product \cite{zhang2015MassiveMIMO}, and the concatenation \cite{he2015corr} of the matrices. Data fusion is guided by the latest mathematical research \cite[Chapter 7]{qiu2015smart}. \\
5) Only eigenvalues are used for further analyses, while the eigenvectors are omitted. This leads to a much faster data-processing speed and less required memory space. Although some information is lost,  there is still  rich information contained in the eigenvalues \cite{ipsen2014weak}, especially those outliers \cite{benaych2013outliers, tao2013outliers}.\\
6) Particularly,  for a certain RMM, various forms of LES, in the form of \Equs{\VLES{F}{}}{\sum\nolimits_{i=1}^{N}{\varphi_F \left( {{\lambda }_{\mathbf{M},i}} \right)}}, can be constructed  by designing test functions $\varphi_F \left(\cdot \right)$ without  introducing any system error.
Each LES, similar to a filter, provides a unique view-angle. As a result, the system is  understood piece by piece. With a proper LES, we can trace some specific signal.\\

Section \ref{section: background} gives the  mathematical background and theoretical foundation. Spectrum test is introduced as a novel tool.
Section \ref{section: method} studies the details about the RMT-based method.
Section \ref{section: case1} and Section \ref{section: case2}, using the simulated data and field data respectively, study the function designing based on the proposed method.
Section \ref{section: concl} concludes this paper.


\section{Mathematical Background and Theoretical Foundation}
\label{section: background}


\subsection{Random Matrix Modeling}
\label {sect:RandomMatrixModeling}
Operating in a balance situation, power grids  obey
\begin{equation}
\label{Eq:Balance}
\left\{ \begin{aligned}
  & \Delta {{P}_{i}}={{P}_{is}}-{{P}_{i}}\left( \mathbf{V},\boldsymbol{\theta } \right) \\
 & \Delta {{Q}_{i}}={{Q}_{is}}-{{Q}_{i}}\left( \mathbf{V},\boldsymbol{\theta } \right) \\
\end{aligned} \right.,
\end{equation}
where ${P}_{is}$ and ${Q}_{is}$ are the power injections of node $i$, and ${P}_{i}\left( \mathbf{V},\boldsymbol{\theta } \right)$ and ${Q}_{i}\left( \mathbf{V},\boldsymbol{\theta } \right)$ are the power injections of the network, satisfying
\begin{equation}
\label{Eq:PiQi}
\left\{ \begin{aligned}
  & {{P}_{i}}={{V}_{i}}\sum\limits_{j=1}^{n}{{{V}_{j}}\left( {{G}_{ij}}\cos {{\theta }_{ij}}+{{B}_{ij}}\sin {{\theta }_{ij}} \right)} \\
 & {{Q}_{i}}={{V}_{i}}\sum\limits_{j=1}^{n}{{{V}_{j}}\left( {{G}_{ij}}\sin {{\theta }_{ij}}-{{B}_{ij}}\cos {{\theta }_{ij}} \right)} \\
\end{aligned} \right..
\end{equation}

Combining \eqref{Eq:Balance} and \eqref{Eq:PiQi}, we obtain
\begin{equation}
\label{Eq:WXY}
{{\mathbf{w}}_{0}}=f\left( {{\mathbf{x}}_{0}},{{\mathbf{y}}_{0}} \right),
\end{equation}
where ${\mathbf{w}}_{0}$ is the vector of nodes' power injections depending on ${P}_{is}$, ${Q}_{is}$,   ${\mathbf{x}}_{0}$ is the system status variables depending on ${V}_{i}$, ${\theta}_{i}$, and  ${\mathbf{y}}_{0}$ is the network topology parameters depending on ${B}_{ij}$, ${G}_{ij}$.

Then, the system fluctuations, thus randomness in datasets, are formulated as
\begin{equation}
\label{Eq:WXYDelta}
{{\mathbf{w}}_{0}}\!+\!\Delta \mathbf{w}\!=\!f\left( {{\mathbf{x}}_{0}}\!+\!\Delta \mathbf{x},{{\mathbf{y}}_{0}}\!+\!\Delta \mathbf{y} \right).
\end{equation}
With a Taylor expansion,~\eqref{Eq:WXYDelta} is rewitten as
\begin{equation}
\label{Eq:WXYDeltaTaylor}
\begin{aligned}
   {{\mathbf{w}}_{\!0}}\!+\!\Delta\! \mathbf{w}\!=&\!f\!\left( {{\mathbf{x}}_{\!0}},\!{{\mathbf{y}}_{\!0}} \!\right)\!+\!f{{'}_{\!\mathbf{x}}}\!\left( {{\mathbf{x}}_{\!0}},\!{{\mathbf{y}}_{\!0}} \!\right)\Delta\! \mathbf{x}\!+\!f{{'}_{\!\mathbf{y}}}\!\left( {{\mathbf{x}}_{\!0}},\!{{\mathbf{y}}_{\!0}} \!\right)\Delta\! \mathbf{y} \\
 & \!+\!\frac{1}{2}f'{{'}_{\!\mathbf{x\!x}}}\!\left( {{\mathbf{x}}_{\!0}},\!{{\mathbf{y}}_{\!0}} \!\right){{\!\left( \Delta\! \mathbf{x} \!\right)}^{\text{2}}}\!+\frac{1}{2}\!f'{{'}_{\!\mathbf{y\!y}}}\!\left( {{\mathbf{x}}_{\!0}},\!{{\mathbf{y}}_{\!0}} \!\right){{\!\left( \Delta\! \mathbf{y} \!\right)}^{\text{2}}} \\& +f'{{'}_{\!\mathbf{x\!y}}}\!\left( {{\mathbf{x}}_{\!0}},\!{{\mathbf{y}}_{\!0}} \!\right)\Delta\! \mathbf{x}\Delta\! \mathbf{y}\!\!+\!\cdots   .  \\
\end{aligned}
\end{equation}

The value of system status variables $\mathbf{x}$ are relatively stable, which means we can ignore the second-order term ${{\left( \Delta \mathbf{x} \right)}^{2}}$ and higher-order terms.
Besides, \eqref{Eq:PiQi} shows that $f'{{'}_{\!\mathbf{y\!y}}}\!\left(\! \mathbf{x},\!\mathbf{y} \!\right)\!=\!0$.
As a result, \eqref{Eq:WXYDeltaTaylor} is turned  into
\begin{equation}
\label{Eq:WXYDeltaTaylorSimplify}
\begin{aligned}
  \Delta\! \mathbf{w}\!=&\!f{{'}_{\!\mathbf{x}}}\!\left( {{\mathbf{x}}_{\!0}},\!{{\mathbf{y}}_{\!0}} \!\right)\Delta\! \mathbf{x}\!+\!f{{'}_{\!\mathbf{y}}}\!\left( {{\mathbf{x}}_{\!0}},\!{{\mathbf{y}}_{\!0}} \!\right)\Delta\! \mathbf{y}   \\&\!+\!f'{{'}_{\!\mathbf{x\!y}}}\!\left( {{\mathbf{x}}_{\!0}},\!{{\mathbf{y}}_{\!0}} \!\right)\Delta\! \mathbf{x}\Delta\! \mathbf{y}\!.
\end{aligned}
\end{equation}

Suppose the network topology is unchanged, i.e., $\!\Delta\!\mathbf{y}\!=\!0$. From~\eqref{Eq:WXYDeltaTaylorSimplify}, we deduce that
\begin{equation}
\label{Eq:Y_0}
\Delta \mathbf{x}\!=\!{{\left(f{{'}_{\!\mathbf{x}}}\!\left(\! {{\mathbf{x}}_{\!0}},\!{{\mathbf{y}}_{\!0}}\! \right) \right)}^{-1}}\!\left(\! \Delta\! \mathbf{w} \! \right)\!=\!{{\mathbf{S}}_{0}}\Delta \mathbf{w}.
\end{equation}

On the other hand, suppose the power demands is unchanged, i.e., $\!\Delta\!\mathbf{w}\!=\!0$. From \eqref{Eq:WXYDeltaTaylorSimplify}, we deduce that
\begin{equation}
\label{Eq:W_0}
\Delta \mathbf{x}\!=\!{{\mathbf{S}}_{0}}\Delta \mathbf{w}_y,
\end{equation}
where $\mathbf{w}_y\!=\![\mathbf{I}\!+\!f'{{'}_{\!\mathbf{x\!y}}}\!\left( {{\mathbf{x}}_{\!0}},\!{{\mathbf{y}}_{\!0}} \!\right)\!\Delta\!\mathbf{y}\!\mathbf{s}_0]^{\!-\!1}
[\-\!f{{'}_{\!\mathbf{y}}}\!\left( {{\mathbf{x}}_{\!0}},\!{{\mathbf{y}}_{\!0}} \!\right)]     .$

Note that $\mathbf{S}_{0}\!=\!\!{{\left(f{{'}_{\!\mathbf{x}}}\!\left(\! {{\mathbf{x}}_{\!0}},\!{{\mathbf{y}}_{\!0}}\! \right) \right)}^{-1}}$, i.e., the inversion of the Jacobian matrix $\mathbf{J}_0$.

Thus, we describe the power system operation using a random vector. If there exists an unexpected active power change or short circuit,  the corresponding change of system status variables ${\mathbf{x}}_{0}$, i.e. ${V}_{i}$, ${\theta}_{i}$, will obey  \eqref{Eq:Y_0} or \eqref{Eq:W_0} respectively.

For a practical system without dramatic changes,
rich statistical empirical evidence indicates that the Jacobian
matrix $\mathbf{J}$ keeps nearly constant, so does $\mathbf{s}_{0}.$ Considering $T$ random vectors observed at time instants $i=1,\cdots,T,$   we build a relationship in the form of
$\Delta \mathbf{X}_\mathbf{s}\!=\!\mathbf{S}_0 \Delta \mathbf{W}$ with a similar procedure as \eqref{Eq:WXY} to \eqref{Eq:W_0}, where $\Delta \mathbf{X}_\mathbf{s}$ denotes the variation of state  $\left[\Delta \mathbf{x}_1, \cdots, \Delta \mathbf{x}_T\right],$ and  $\Delta \mathbf{W}$ denotes the variation of power injections or topology parameters accordingly.

Taking the case in \cite{he2015les} as an example, for an equilibrium operation system (the topology is unchanged, the reactive power is almost constant or changes much more slowly than the active one),  the relationship model between  voltage magnitude and active power is  just like the Multiple Input Multiple Output (MIMO) model in wireless communication~\cite{qiu2015smart,zhang2015MassiveMIMO}. We write  $ \mathbb{V} = {\mathbf{\Xi}} {\mathbb{P}}$. Note that most variables of vector $\mathbb{V}$ are random due to the ubiquitous noises, e.g., small random fluctuations in $\mathbb{P}$. Furthermore, with the normalization, we can build the standard random matrix model (RMM) in the form of $\tilde{\mathbb{V}} = \tilde{\mathbf{\Xi}}\mathbf{R}$, where  $\mathbf{R}$ is a standard Gaussian random matrix.

\subsection{Anomaly Detection Based on Asymptotic Empirical Spectral Distribution}
Often, these rapid fluctuations exhibit Gaussian statistical properties~\cite {lim2016svd}, as pointed out above. In practice, Gaussian unitary ensemble (GUE) and Laguerre unitary ensemble (LUE) are used  in our models:
\begin{equation}
\label{eq:B1}
\mathbf{A}=\left\{ \begin{aligned}
  & \frac{1}{2}\left( \mathbf{X}+{{\mathbf{X}}^{H}} \right)&, \mathbf{X}\in {{\mathbb{X}}^{N\times N}}& \text{  ,GUE;} \\
 & \frac{1}{N}\mathbf{X}{{\mathbf{X}}^{H}}&, \mathbf{X}\in {{\mathbb{X}}^{N\times T}}& \text{  ,LUE}\text{.} \\
\end{aligned} \right.,
\end{equation}
where $\mathbf{X}$ is the standard Gaussian random matrix whose entries are independent
identically distributed (i.i.d.) complex Gaussian random variables.

Let ${f_{\bf{A}}}\left( x \right)$ be the empirical density of $\bf{A}$, and define its empirical spectral distribution (ESD) ${F_{\bf{A}}}\left( x \right)$:
\begin{equation}
\label{eq:D1}
{F_{\bf{A}}}\left( x \right) = \frac{1}{N}\sum\limits_{i = 1}^N {{I_{\left\{ {{\lambda _i} \le x} \right\}}}},
\end{equation}
where $\bf{A}$ is GUE or LUE matrix, and $I\left(  \cdot  \right)$ represents the event indicator function. We investigate the rate of convergence of the expected ESD $\mathbb{E}\left\{ {{F_{\bf{A}}}\left( x \right)} \right\}$ to the Wigner's Semicircle Law or Wishart's M-P Law.

Let ${g_{\bf{A}}}\left( x \right)$ and ${G_{\bf{A}}}\left( x \right)$ denote the true eigenvalue density and the true spectral distribution of $\bf{A}$, and the Wigner's Semicircle Law and Wishart's M-P Law say:

\begin{equation}
\label{eq:D2}
{g_{\bf{A}}}\left( x \right) = \left\{ \begin{aligned}
&\frac{1}{{2\pi }}\sqrt {4 - {x^2}} &, x \in \left[ { - 2,2} \right]&\text{  ,GUE}; \\
&\frac{1}{{2\pi cx}}\sqrt {\left( {x - a} \right)\left( {b - x} \right)} &, x \in \left[ { a,b} \right] &\text{  ,LUE};
\end{aligned} \right.,
\end{equation}
where $a = {\left( {1 - \sqrt c  } \right)^2}, b = {\left( {1 + \sqrt c  } \right)^2}$.

\begin{equation}
\label{eq:DD}
{G_{\bf{A}}}\left( x \right) = \int_{ - \infty }^x {{g_{\bf{A}}}\left( u \right)du} .
\end{equation}
Then, we denote the Kolmogorov distance between $\mathbb{E}\left\{ {{F_{\bf{A}}}\left( x \right)} \right\}$ and ${G_{\bf{A}}}\left( x \right)$ as  $\Delta$:
\begin{equation}
\label{eq:D3}
\Delta  = \mathop {\operatorname{sup}}\limits_x \left| {\mathbb{E}\left\{ {{F_{\bf{A}}}\left( x \right)} \right\} - {G_{\bf{A}}}\left( x \right)} \right|.
\end{equation}

Gotze and Tikhomirov, in their work \cite{gotze2005rate}, prove an optimal bound for $\Delta$  of order $O\left( {{N^{ - 1}}} \right)$.

\newtheorem{thm1}{Lemma}[section]
\begin{thm1}
\label{thm1:Convergence for Spectra}
There exists a positive constant $C$ such that, for any $N \geq 1 $,
\begin{equation}
\label{eq2:Convergence for Spectra}
\Delta  \le C{N^{ - 1}}.
\end{equation}
\end{thm1}

They also prove that the convergence of the density of standard Semicircle Law and M-P Law to the expected spectral density $f_{\bf{A}}(x)$ satisfies following lemmas.
\newtheorem{thm2}[thm1]{Lemma}
\begin{thm2}
\label{thm2:Convergence for Spectra}
For {\rm{GUE}} matrix, there exists a positive constant $\varepsilon$ and $C$ such that, for any $x \in \left[ { - 2 + {N^{ - \frac{1}{3}}}\varepsilon ,2 - {N^{ - \frac{1}{3}}}\varepsilon } \right],$
\begin{equation}
\label{eq3:Convergence for Spectra}
\left| {f_{\bf{A}}\left( x \right) - g\left( x \right)} \right| \le \frac{C}{{N\left( {4 - {x^2}} \right)}}.
\end{equation}
\end{thm2}

\newtheorem{thm5}[thm1]{Lemma}
\begin{thm5}
\label{thm5:Convergence for Spectra}
For {\rm{LUE}} matrix, let $\beta  ={N}/{T}$, there exists some positive constant $\beta _1$ and $\beta _2$ such that $0 < {\beta _1} \le \beta  \le {\beta _2} < 1$, for all $N \geq 1 $. Then there exists a positive constant $C$ and $\varepsilon$ depending on $\beta _1$ and $\beta _2$ and  for any $N \geq 1 $ and $x \in \left[ {a + {N^{ - \frac{2}{3}}}\varepsilon ,b - {N^{ - \frac{2}{3}}}\varepsilon } \right],$
\begin{equation}
\label{eq8:Convergence for Spectra}
\left| {f_{\bf{A}}\left( x \right) - h\left( x \right)} \right| \le \frac{C}{{N\left( {x - a} \right)\left( {b - x} \right)}}.
\end{equation}
\end{thm5}

Lemma \ref{thm2:Convergence for Spectra} and \ref{thm5:Convergence for Spectra} also describe how fast the population distribution functions converge to the asymptotic empirical spectral distribution limit. This  ESD-based test is interesting for anomaly detection about a complex grid; the effectiveness is validated  in Section \ref{section: case1}. We exploit the mathematical knowledge that the ESD converges to its limit  with a  optimal convergence rate of ${N}^{-1}.$

\section{The Method of Situation Awareness}
\label{section: method}

\subsection{Technical Route and Practical Procedures}
\begin{figure}[htbp]
\centerline{
\includegraphics[width=.37\textheight,height=.26\textwidth]{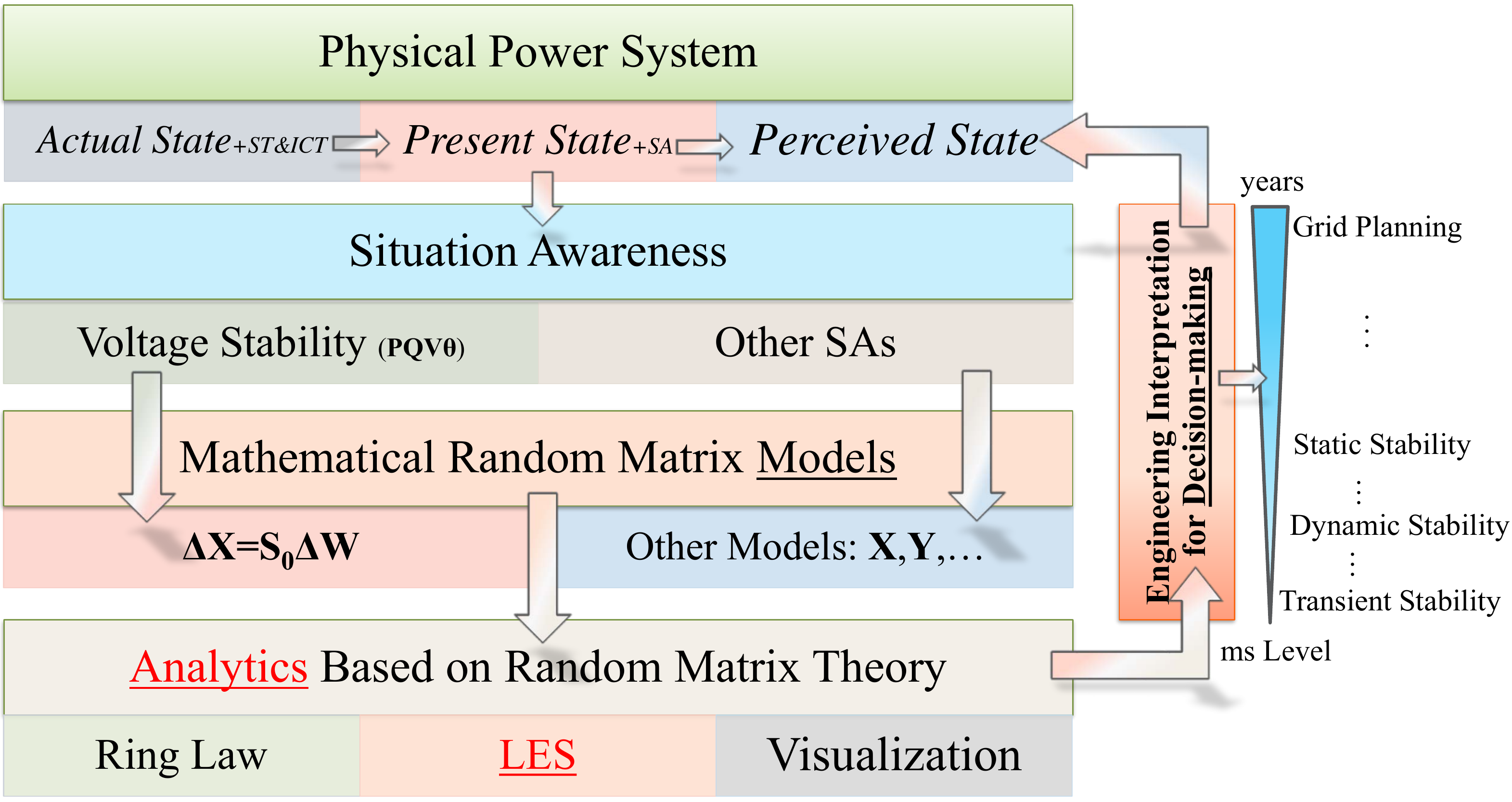}
}
\caption{RMT-based Method for SA}
\label{fig:methodSA}
\end{figure}

The proposed RMT-based method consists of  three procedures as illustrated in Fig. \ref{fig:methodSA}: 1) big data model---to model the system using experimental data for the RMM; 2) big data analysis---to conduct big data anlytics for the indicator system; 3) engineering interpretation---to visualize and interpret the statistical results to operators for the decision-making.

 This method is universal. We have already made numerous successful attempts in the field of anomaly detection and diagnosis for both the grid network \cite{he2015arch, he2015corr} and the transmission equipment \cite{yan2016transmission}. In addition, Zhang et al., based on RMT, study the steady stability and transient stability in \cite{Wu2016A} and \cite{Liu2016Power} respectively.

\subsection{Paradigms and Method}

\begin{figure}[htbp]
\centerline{
\includegraphics[width=.3\textheight,height=.3\textwidth]{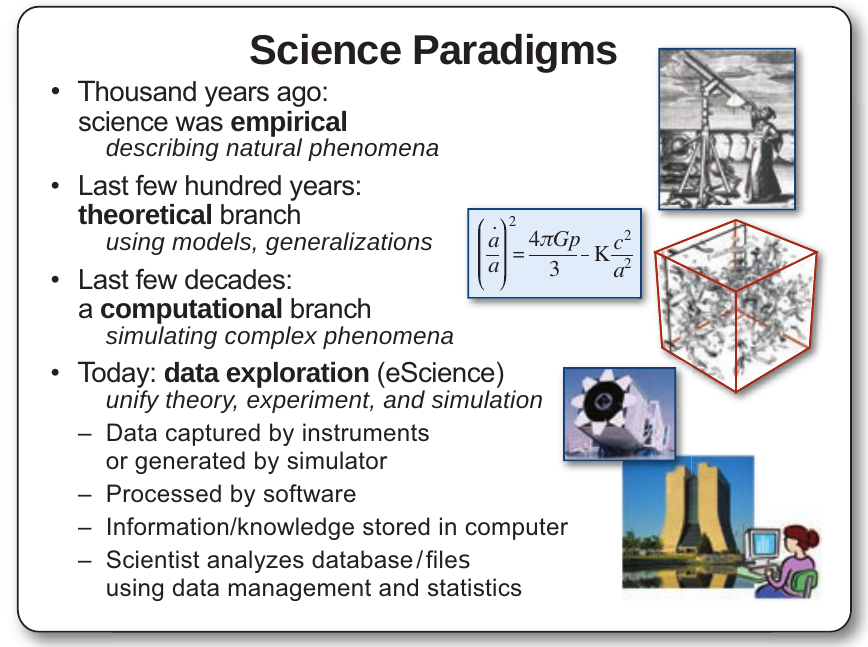}
}
\caption{Science paradigms \cite{gray2009jim}}
\label{fig:fourthp}
\end{figure}

\begin{figure*}[htb]
\includegraphics[width=.72\textheight,height=.7\textwidth]{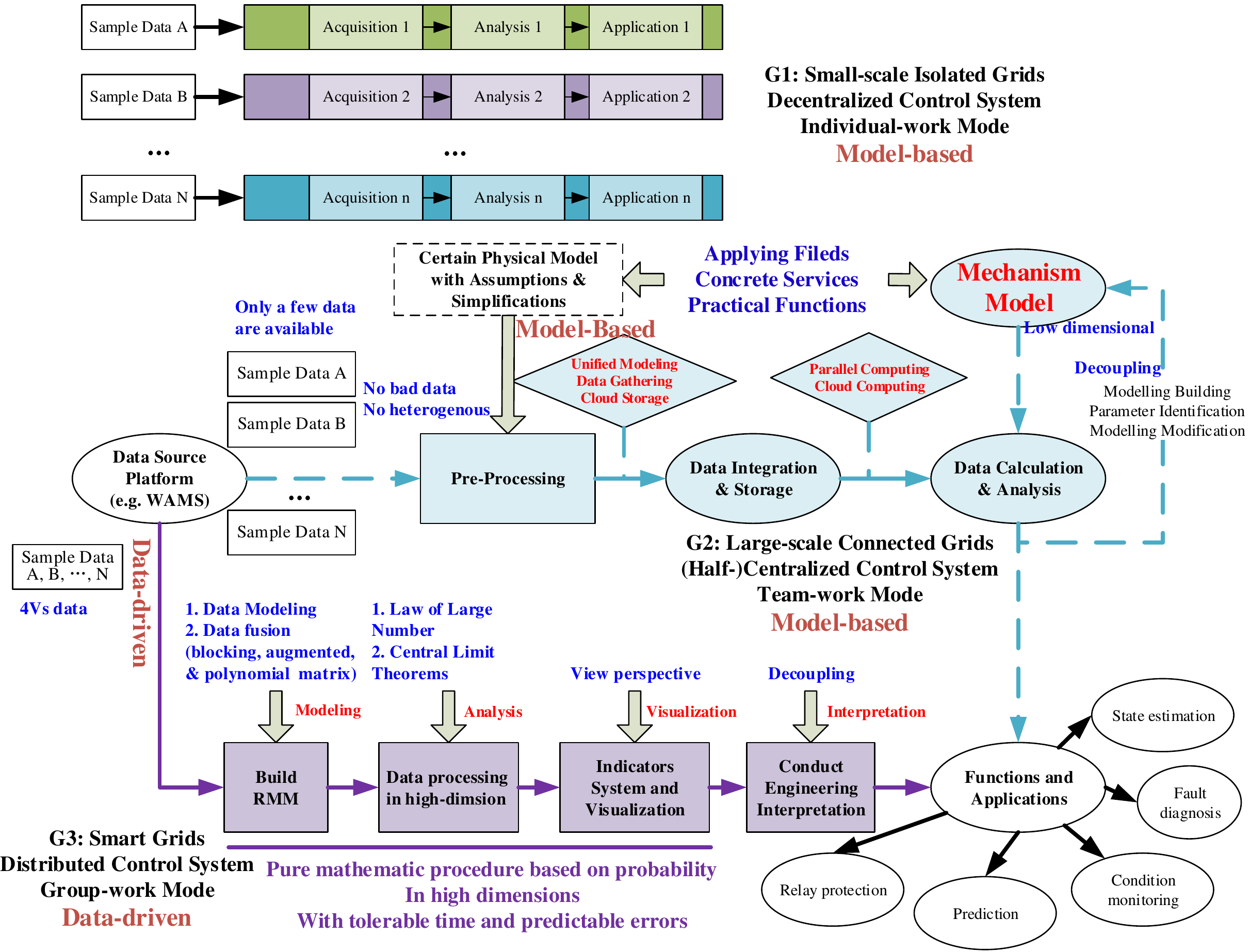}
\caption{Data utilization  method for power systems. The above, middle, and below parts indicate the data processing procedures and the work modes for G1, G2, and G3, respectively.}
\label{fig:procedure}
\end{figure*}

We would like to refer to Fig. \ref{fig:fourthp} in book \cite{hey2009fourth} as a clue. We are now entering the age of  4th-paradigm---data-intensive scientific discovery.
Besides, the summaries for the  classical decision-making approaches and for our proposed ones, obtained initially in \cite{he2015arch}, are improved as Fig. \ref{fig:procedure}.

The second and third paradigms are typically model-based---they use equations, formulas, and simulations to describe the system.
The blue line in Fig. \ref{fig:procedure} depicts the general procedure and corresponding tools.
These tools cannot deal with massive data due to the essence of mechanism models---the models are in low dimensions, leading to deterministic results which are fully dependent upon only a few parameters\footnote{E.g., $y\!=\!ax^2\!+\!bx\!+\!c$ is a 3-dimensional model---the relationship between $x$ and $y$ fully depends on $a$, $b$, and $c$.}. It will raise some problems, causing inefficient or even incorrect big data analytics. For instance, only under ideal conditions, is the wind power proportional to the cube of wind speed. Moreover, some physical parameters, e.g., admittance matrixes, will introduce system error due to the ubiquitous randomness and uncertainty.

Under classical statistical framework, only two typical data matrices in the form of $\mathbf{X}\!\in\! {\mathbb{R}}^{N\times T}$ are at our disposal:  1)  $N,T$ are small, and 2) $N$ is small, $T$ is very large (compare with $N$).
This prerequisite greatly restricts the utilization of the massive data; we should enable more data to speak for themselves \cite{kitchin2013big}.
In other words, model-based framework is not able to turn massive data into useful big data analytics. Although these massive data can contribute to model improvement and parameters correction, we can hardly conduct analysis more precisely with extremely large data volumes. Even worse, more data mean more errors; if we take those bad data into the fixed model, poor results are obtained almost surely.
Besides, the bias, caused by challenges such as error accumulations and spurious correlations, will not be alleviated via a low-dimensional procedure \cite{he2015corr}---the dimensions of the procedure are limited by the dimensions of the model.
The belief that data-driven mode is adapted to the future  grid's analysis agrees with  the core viewpoint of the 4th-paradigm. The classical data utilization methodology needs be revisited.

\subsection{Classical Dimensionality Reduction Algorithm---PCA}
Data-driven methodology is an alternative; it is model-free and able to process massive data in a holistic way.
Principal component analysis (PCA) is one of the classical data processing algorithms which are sensitive to relative scaling original variables. It uses an orthogonal transformation to convert a set of  possibly correlated raw variables into a set of linearly uncorrelated variables called principal components. The number of principal components is often much less than the number of original variables. In \cite{xie2014dimensionality}, PCA is used  for  dimensionality reduction  from 14 PMU datasets to extract the event indicators. For PCA, the procedure consists of three parts:  1) Singular Value Decomposition (SVD) \cite{lim2016svd}, 2) Projection, and 3) Indicators.

This procedure is applied to conduct early event detection; details can be found in \cite{xie2014dimensionality}.
We will make a comparison between the PCA-based approach and the RMT-based approach, and the advantages of the later is summarized in \ref{AdvantagesofRMT}.

\subsection{Data-Driven Approach Based on Random Matrix Theory}
The procedure based on RMT is outlined below.
\subsubsection{Ring Law and MSR}
\text{\\}
 Ring Law Analysis conducts SA as follows:

\begin{table}[H]
\centering

\begin{tabular*}{8.8cm} {p{8.6cm}}
\toprule[1.5pt]
\textbf {Steps of Ring Law Analysis} \\
\toprule[0.5pt]
1) Select  arbitrary raw data (or all available data) as data source \Vector \Omega.\\
2) At a certain time \Index {t}{i}, form \VRX{} as random matrix.\\
3) Obtain \VTZ{} by matrix transformations (\Data{\VRX{}\rightarrow\VTX{}\rightarrow\VX{u}\rightarrow\VZ{}\rightarrow\VTZ{}} \cite{he2015arch}).\\
4) Calculate eigenvalues \Vlambda{\VTZ{}} and plot the Ring on the complex plane.\\
5) Conduct high-dimensional analysis.\\
\quad 5a) Observe the experimental ring and compare it with the reference.\\
\quad 5b) Calculate \Equs{\VLES{MSR}{}}{\sum\nolimits_{i=1}^{N}{\left| {{\lambda }_{{{\mathbf{Z}},{i}}}} \right|}/N} as the \textit{statistical indicators}.\\
\quad 5c) Compare \VLES{MSR}{} with the theoretical value \Data{}{\STE{\VLES{MSR}{}}}.\\
6) Repeat 2)-5) at the next time point (\Equs {\Index {t}{i}}{\Index {t}{i}+1}).\\
7) Visualize \VLES{}{} on the time series, i.e. draw \VLES{}{}--$t$ curve.\\
8) Make engineering explanations.\\

\toprule[1pt]
\end{tabular*}
\end{table}

In \textit{Steps 2--7}, with a high-dimensional procedure, one conducts SA without any prior knowledge, assumption, or simplification.
In \textit{step 2}, arbitrary raw data, even those from distributed nodes or intermittent time periods, are at our disposal. The size of  \VRX{} is controllable, and as a result the dimensionality curse is  relieved in some ways.

\subsubsection{M-P Law and LES}
For the M-P Law Analysis, the steps are very similar, except for the following differences:
\begin{table}[H]
\centering

\begin{tabular*}{8.8cm} {p{8.6cm}}
\toprule[1.5pt]
\textbf {Partial Steps of M-P Law Analysis} \\
\toprule[0.5pt]
3] Obtain $\mathbf{M}$ by matrix transformations ($\mathbf{M}=\frac{1}{N}\VTX{}{\VTX{}^{H}}$).\\
4] Calculate eigenvalues \Vlambda{\mathbf{M}}.\\
\quad 5b] Calculate \Equs{\VLES{}{}}{\sum\nolimits_{i=1}^{N}{\varphi \left( {{\lambda }_{\mathbf{M},i}} \right)}} as the \textit{statistical indicators}.\\
\quad 5c] Compare \VLES{}{} with the theoretical value \Data{}{\STE{\VLES{}{}}}.\\

\toprule[1pt]
\end{tabular*}
\end{table}
Notice that Ring Law maps the information from datasets to the complex plane (${\mathbb{C}^{N \times T}} \mapsto \mathbb{C}$), while M-P law does this to the right half real-axis (${\mathbb{C}^{N \times T}} \mapsto {\mathbb{R}^ + }$). This fundamental difference plays a critical role in data visualization.

\section{Case Studies Using Simulated Data}
\label{section: case1}

\subsection{Background and Assumption of the Case}
\begin{figure}[htbp]
\subfloat[Assumed event, unavailable.]{\label{fig:Case0Event}
\includegraphics[width=0.22\textwidth]{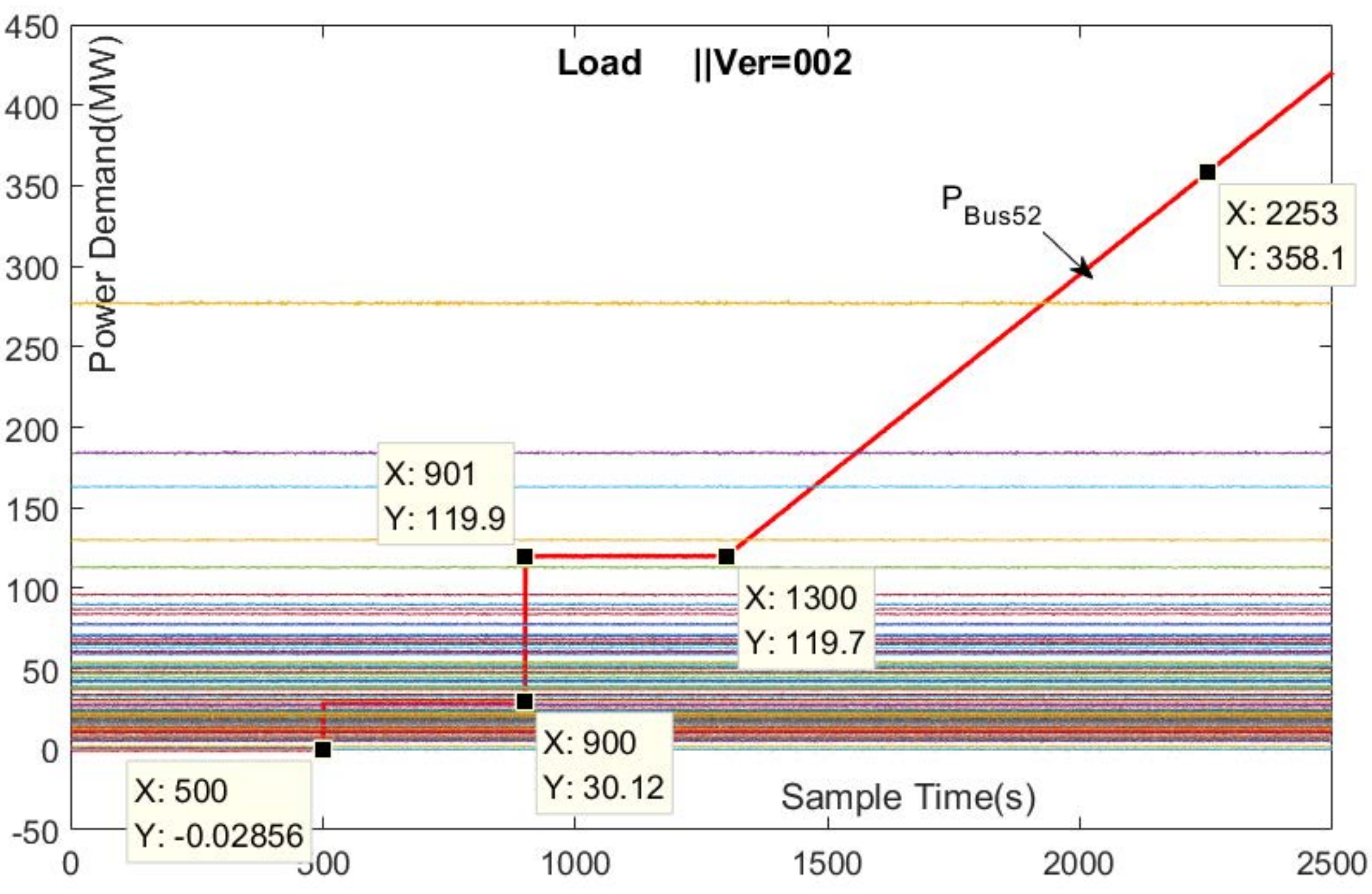}
}
\subfloat[Raw voltage,  $\bm{\Omega}_{\mathbf{V} }$ for analysis.]{\label{fig:Case0Vol}
\includegraphics[width=0.22\textwidth]{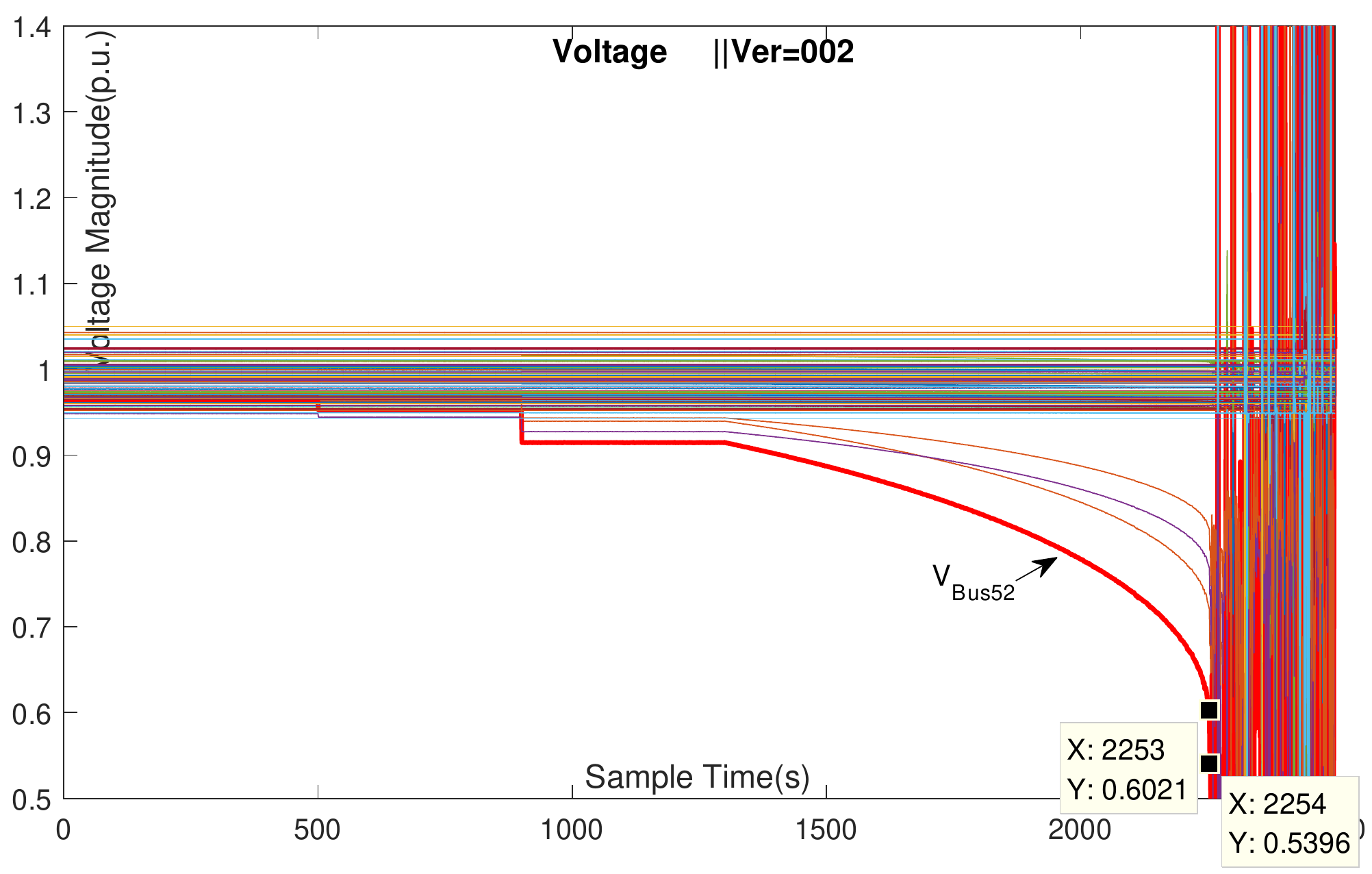}
}

\subfloat[Category for operation status.]{\label{fig:Case0category}
\includegraphics[width=0.46\textwidth]{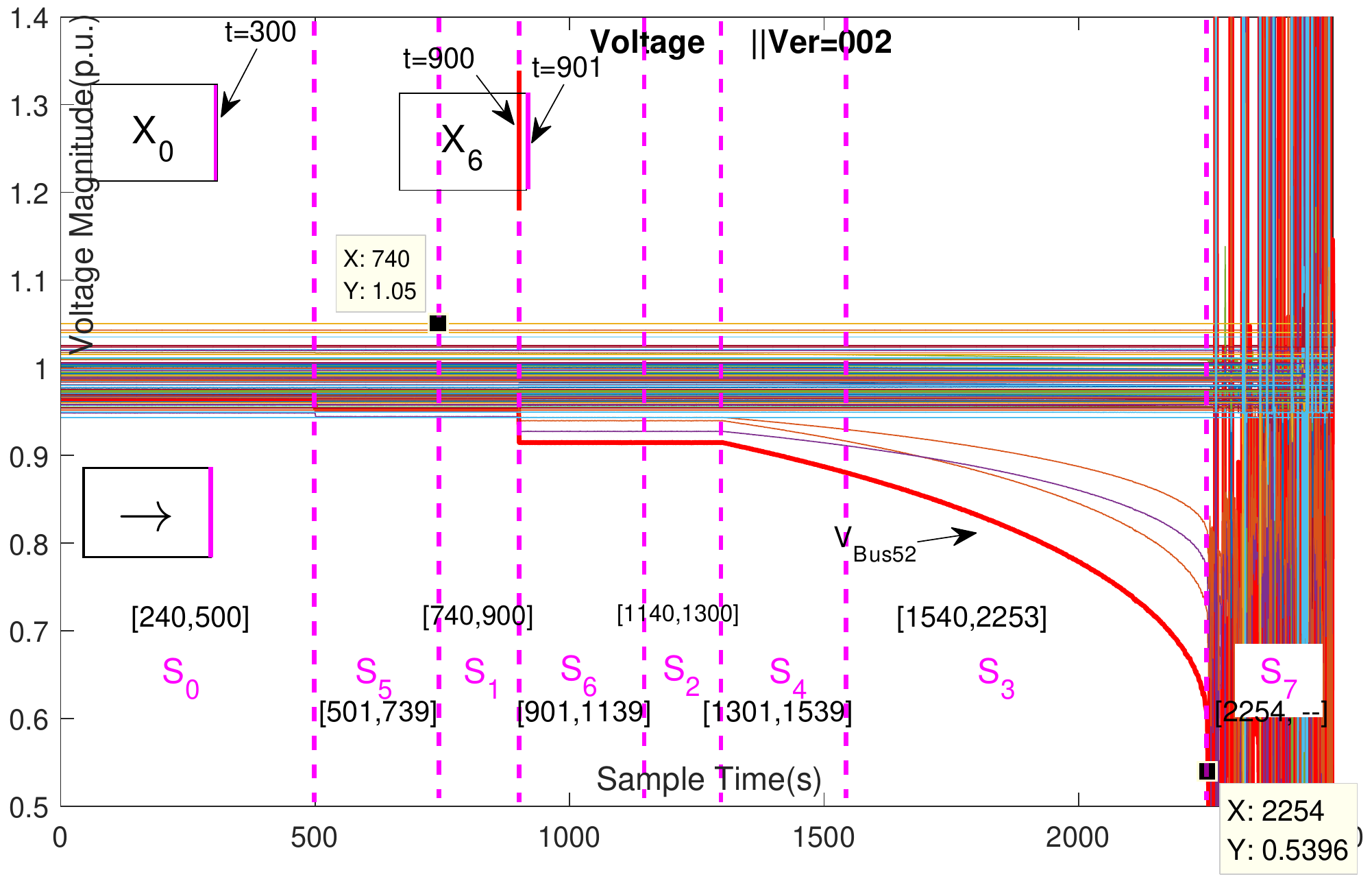}
}

\caption{Background of Case 1.}
\label{fig:Case0A}
\end{figure}

\begin{figure}[H]
\centering
\begin{overpic}[scale=0.58
]{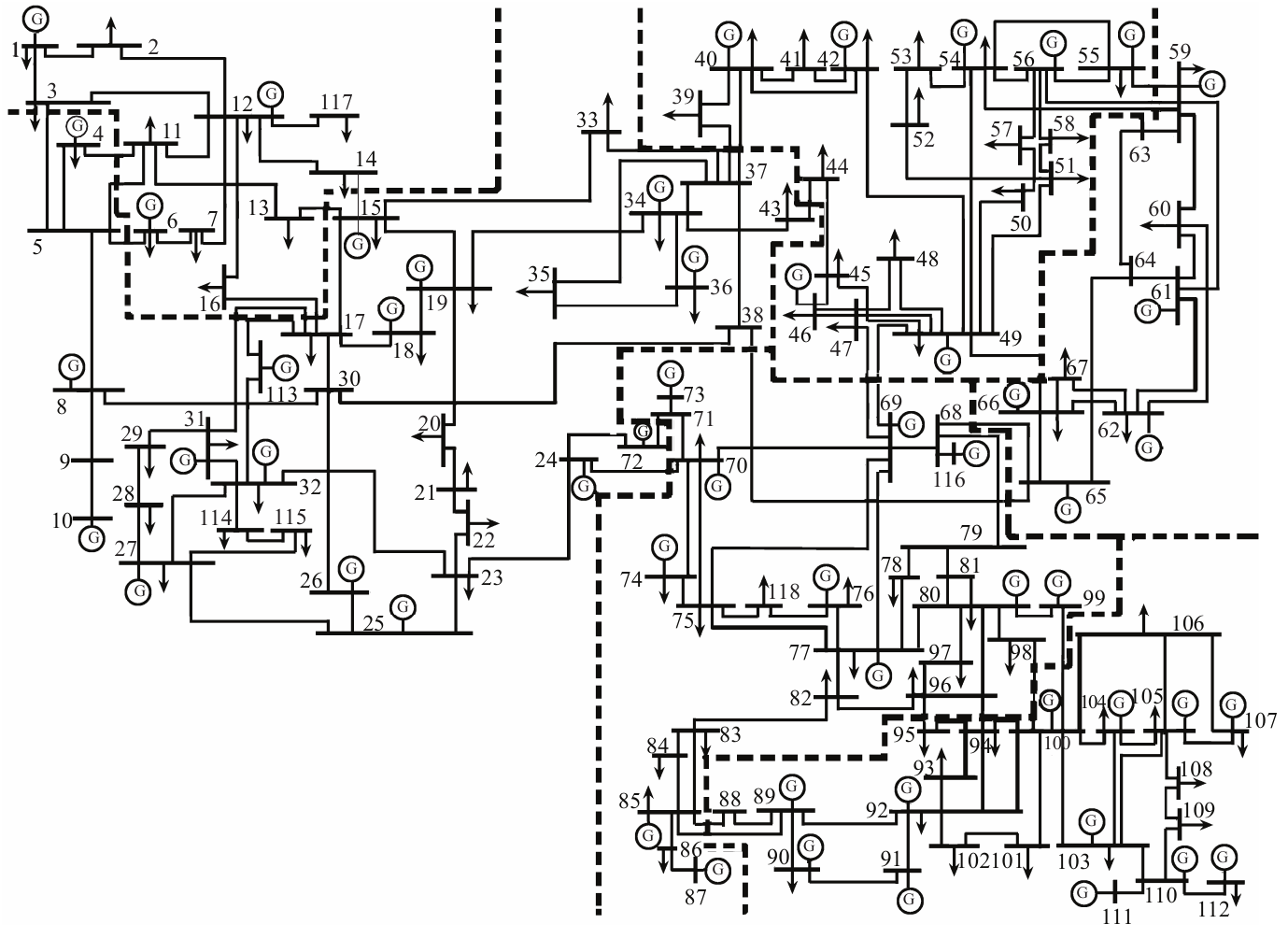}
    \setlength {\fboxsep}{1pt}
    \put(71,60) {\fcolorbox{red}{white}{\tiny \color{blue}{$52$}}} 

      \setlength {\fboxsep}{2pt}

\end{overpic}
\caption{Partitioning network for the IEEE 118-node system.}
\label{fig:IEEE118network}
\end{figure}

We adopt a standard IEEE 118-node system as Fig. \ref{fig:IEEE118network} and assume the events as Table \ref{Tab: Event Series}.
Thus, the power demand on each node is obtained as the system injections (Fig. \ref{fig:Case0Event}); the voltage is also obtained (Fig. \ref{fig:Case0Vol}).
Suppose that the power demand data is \textit{unknown} or unqualified for SA due to the low sampling frequency or the bad quality.
For further analysis, we just start with data source  $\bm{\Omega}_{\mathbf{V} }:{\Rdata v_{i,j}}\Belong{}{\VF R{118}{2500}}$  and assign the analysis matrix as $\mathbf{X}\Belong{}{\VF {R}{118}{240}}$ (4 minutes' time span).
Firstly, we conduct category for the system operation status; the results are given in Fig. \ref{fig:Case0category}.
In general, according to the data feature (events on time-series) and the matrix length (time span, i.e., $T$), we divide the operation satus into 8 stages. Note that $\textbf{S}4, \textbf{S}5$, and $\textbf{S}6$ are transition stages, and their time span is right equal to the analysis matrix  length minus ones, i.e, $T\!-\!1\!=\!239$.
These stages are described as follows:

\begin{itemize}
\item For $\textbf{S}0, \textbf{S}1, \textbf{S}2$, white noises play a dominant part. \VPbus{52} is rising in turn.
\item For $\textbf{S}3$, \VPbus{52} keeps a sustained and stable growth.
\item $\textbf{S}4$, transition stage. Ramping signal exists.
\item $\textbf{S}5, \textbf{S}6$, transition stages. Step signal exists.
\item For $\textbf{S}7$, voltage collapse.
\end{itemize}

Two typical data sections, at stage $\textbf{S}0$ and $\textbf{S}6$ respectively, are selected:  $\Vector X_0\Belong{}{\VF R{118}{240}}$, covering period $t\!=\![61\!:\!300]$ and at sampling time $t_\text{end}\!=\!300$, and 2) $\Vector X_6\Belong{}{\VF R{118}{240}}$, covering period $t\!=\![662\!:\!901]$ and at sampling time $t_\text{end}\!=\!901$.

\subsection{Anomaly Detection}
\subsubsection{Based on Ring Law and M-P Law}
\text{\\}

\begin{figure}[htbp]
\centering
\subfloat[Ring Law for $\Vector X_0$]{\label{fig:Case0X0ring}
\includegraphics[width=0.18\textwidth]{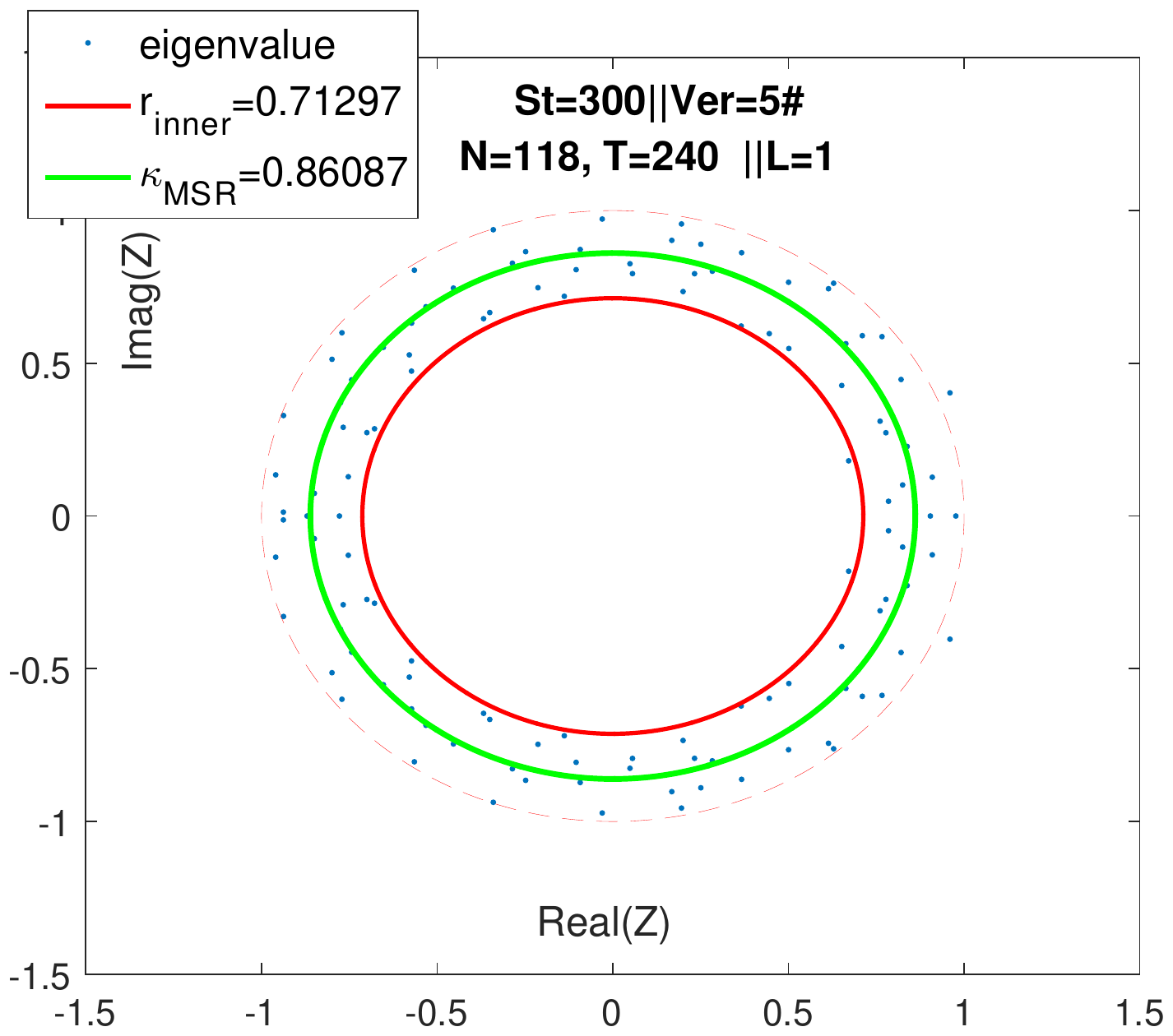}
}
\subfloat[M-P Law  for $\Vector X_0$]{\label{fig:Case0X0mp}
\includegraphics[width=0.24\textwidth]{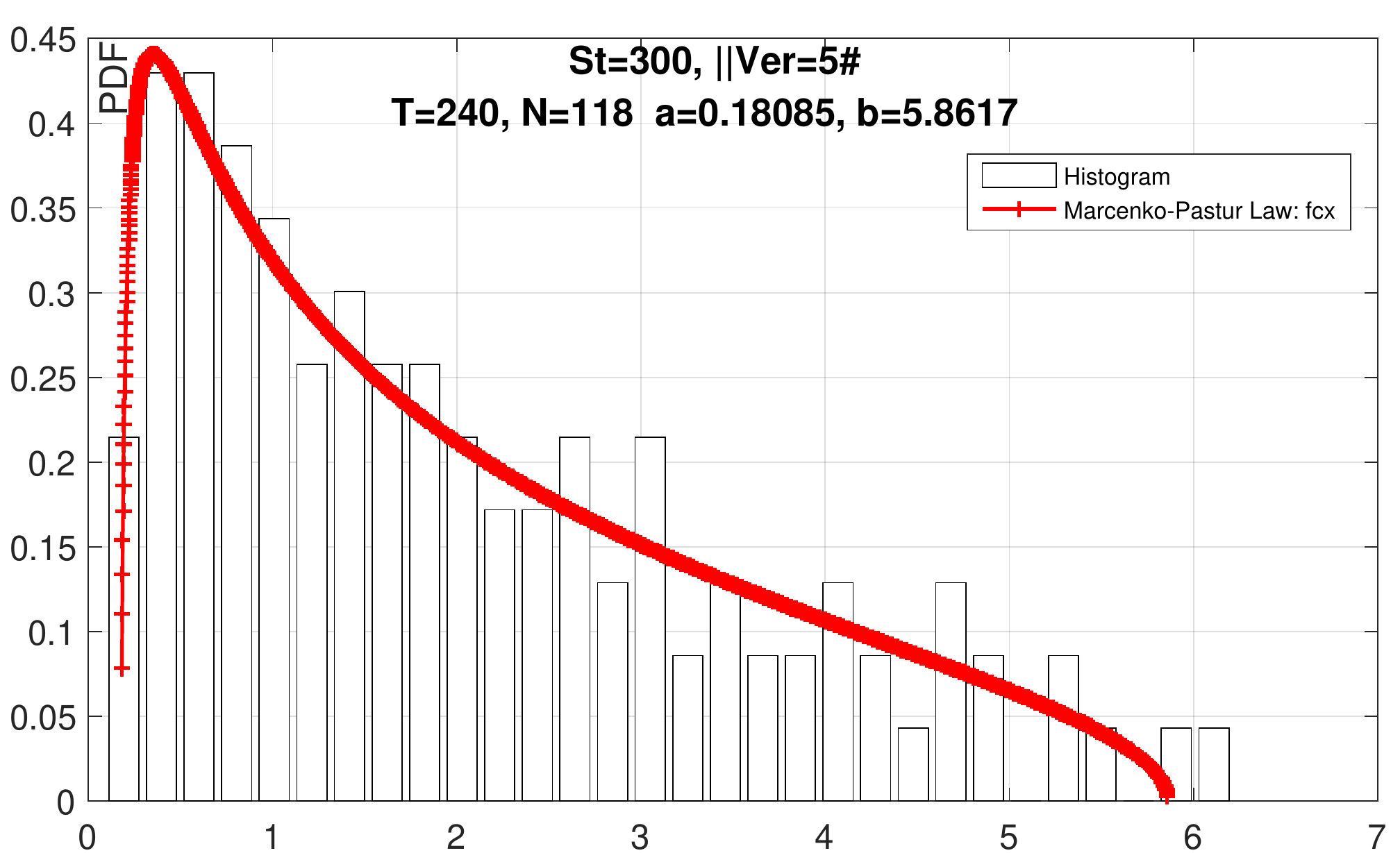}
}

\subfloat[Ring Law  for $\Vector X_6$]{\label{fig:Case0X6ring}
\includegraphics[width=0.18\textwidth]{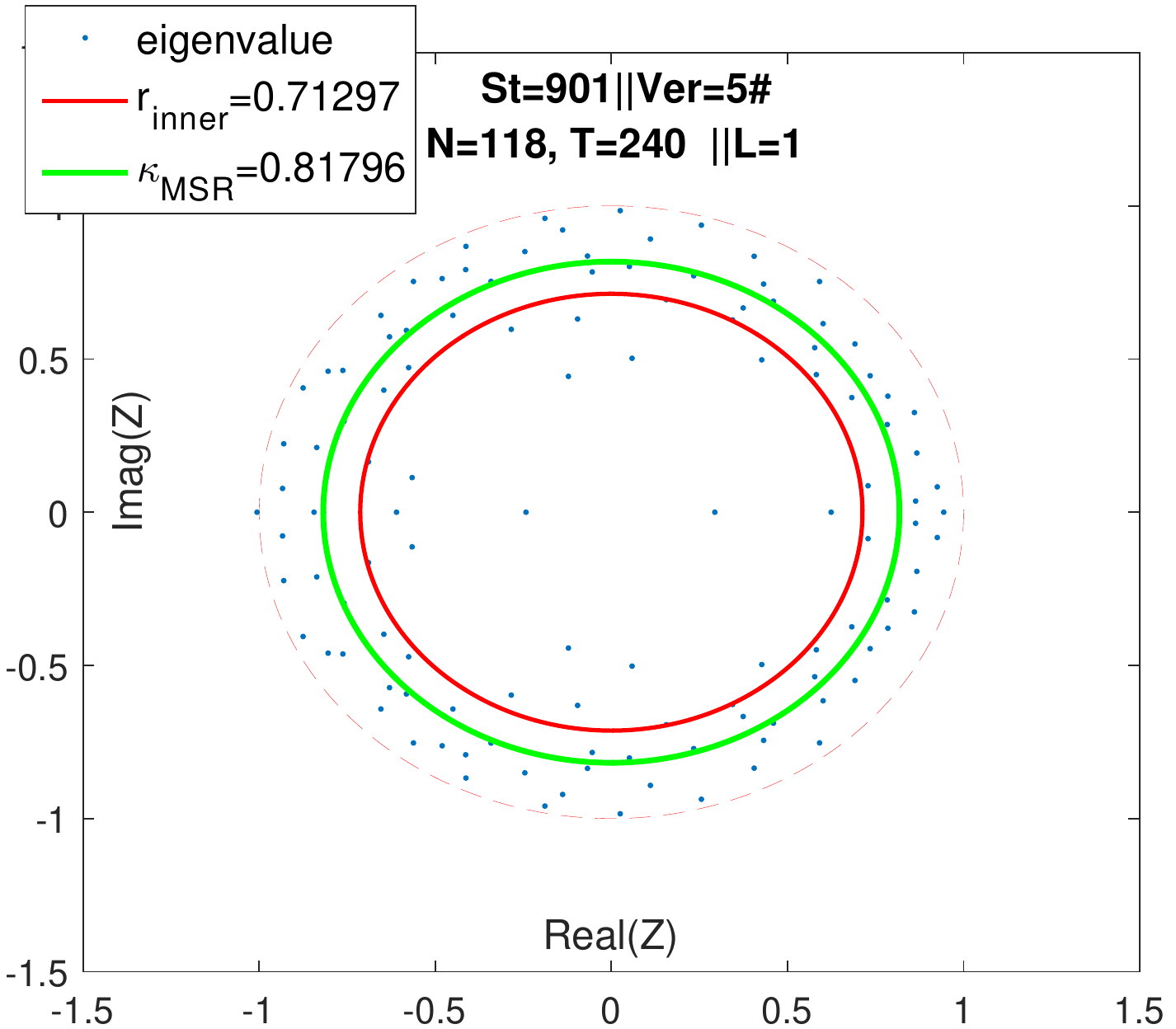}
}
\subfloat[M-P Law  for $\Vector X_6$]{\label{fig:Case0X6mp}
\includegraphics[width=0.24\textwidth]{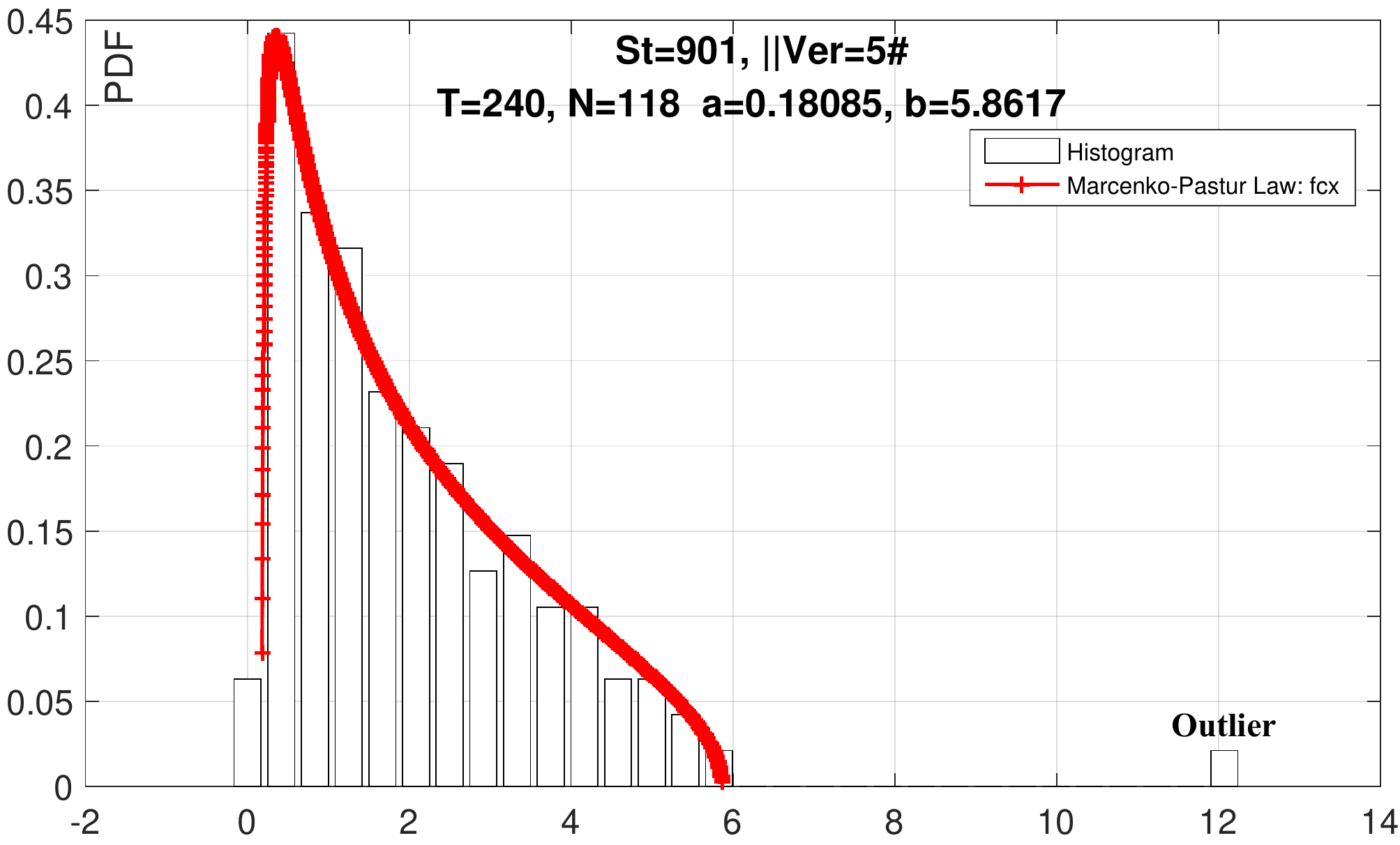}
}

\subfloat[\Cur{\tau_{\text{MSR}}}{t} curve using MSW method on time series.]{\label{fig:Case0msr}
\includegraphics[width=0.46\textwidth]{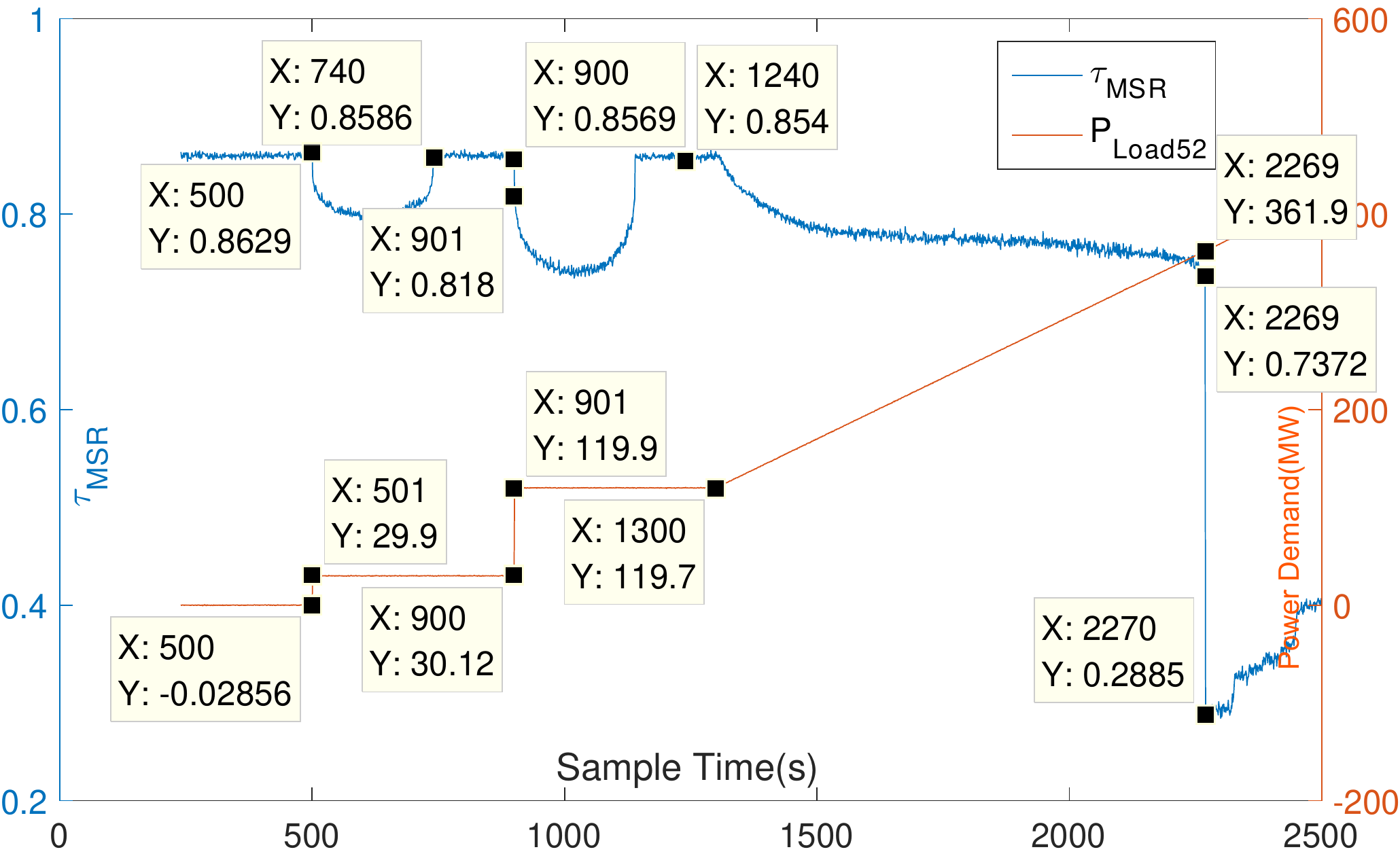}
}
\caption{Anomaly detection results.}
\label{fig:Case00A}
\end{figure}

According to our previous work \cite{he2015arch}, RMM $\tilde{\mathbb{V}}$ is build from the raw voltage data. Then,  $\tau_{\text{MSR}}$ is employed as a statistical indicator to conduct anomaly detection. For the selected  data section $\Vector X_0$ and $\Vector X_6$, their M-P Law and Ring Law Analysis are shown as Fig \ref{fig:Case0X0ring}, \ref{fig:Case0X0mp}, \ref{fig:Case0X6ring} and  \ref{fig:Case0X6mp}. With sliding-window,  the \Cur{\tau_{\text{MSR}}}{t} curve is obtained as Fig. \ref{fig:Case0msr}.
\begin{figure}[htbp]
\centering
\includegraphics[width=0.47\textwidth]{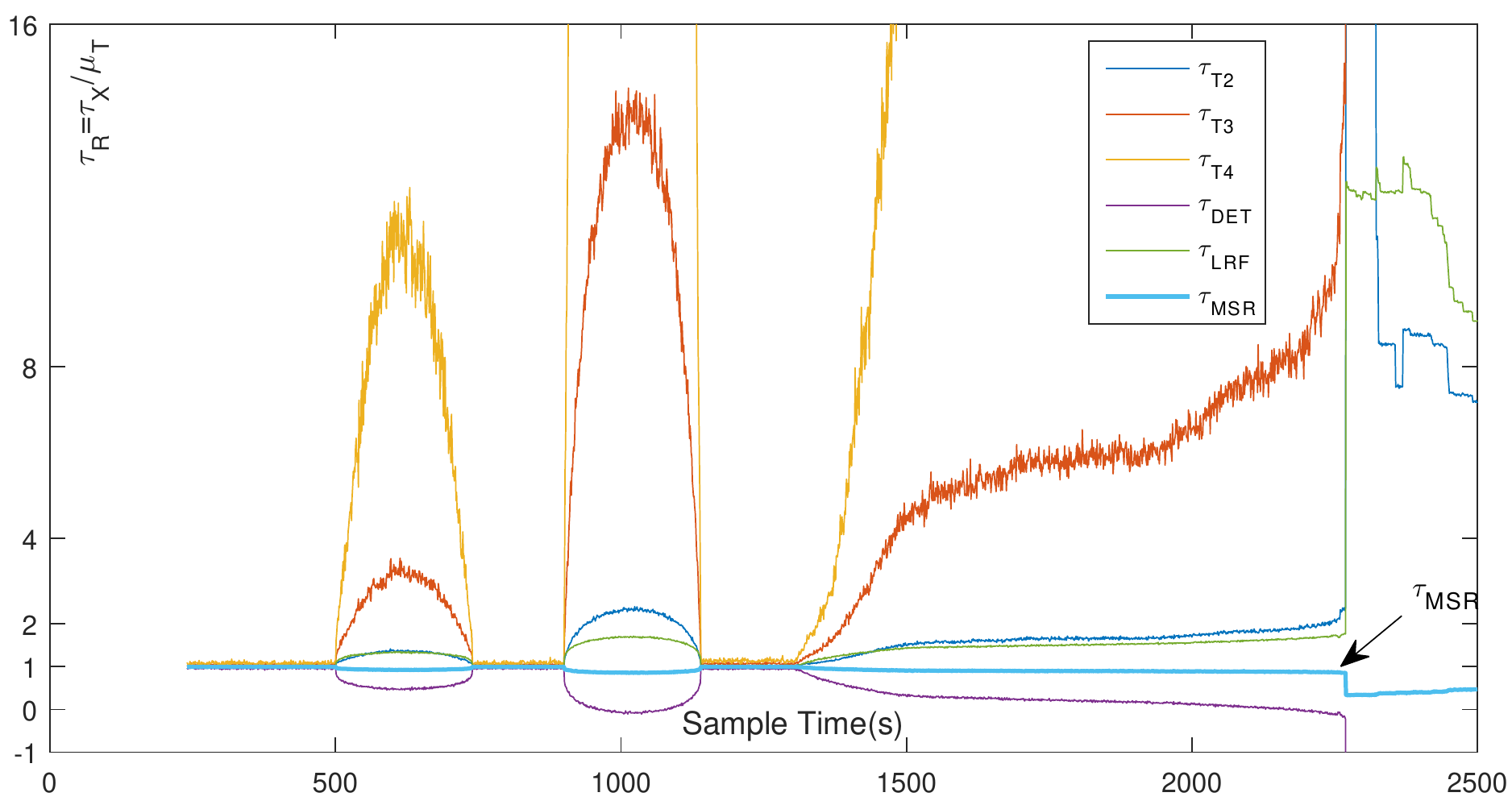}
\caption{Illustration of various LES indicators.}
\label{fig:Case0LESs}
\end{figure}

Fig. \ref{fig:Case00A} shows that  when there is no signal in the system, the experimental RMM well matches Ring Law and M-P Law, and the experimental value of LES is approximately equal to the theoretical value.  This validates the theoretical justification for modeling rapid fluctuation of each node using white Gaussian noises, as shown in Section \ref{sect:RandomMatrixModeling}. On the other hand, Ring Law and M-P Law are violated at the very beginning ($t_\text{end}\!=\!901$) of the step signal.
Besides, the proposed high-dimensional indicator  $\tau_{\text{MSR}}$, is extremely sensitive to the anomaly---at $t_\text{end}\!=\!901$, the $\tau_{\text{MSR}}$ starts the dramatic change (Fig. \ref{fig:Case0msr},  \Cur{\tau_{\text{MSR}}}{t} curve), while the raw voltage magnitudes are still in the normal range (Fig. \ref{fig:Case0category}). 
Moreover, following the previous work \cite{he2015les}, we design numerous kinds of LES $\tau$ and define $\mu_0\!=\!\tau/\STE{\tau}.$ The results are shown in Fig. \ref{fig:Case0LESs}, proving that  different indicators have different effectiveness; this suggests another topic to explore in the future.

\subsubsection{Based on Spectrum Test}
\text{\\}

\begin{figure}[hbtp]
\subfloat[ESD of $\Vector Y_0$ (Normal)]{
\includegraphics[width=0.23\textwidth, height=1.3 in]{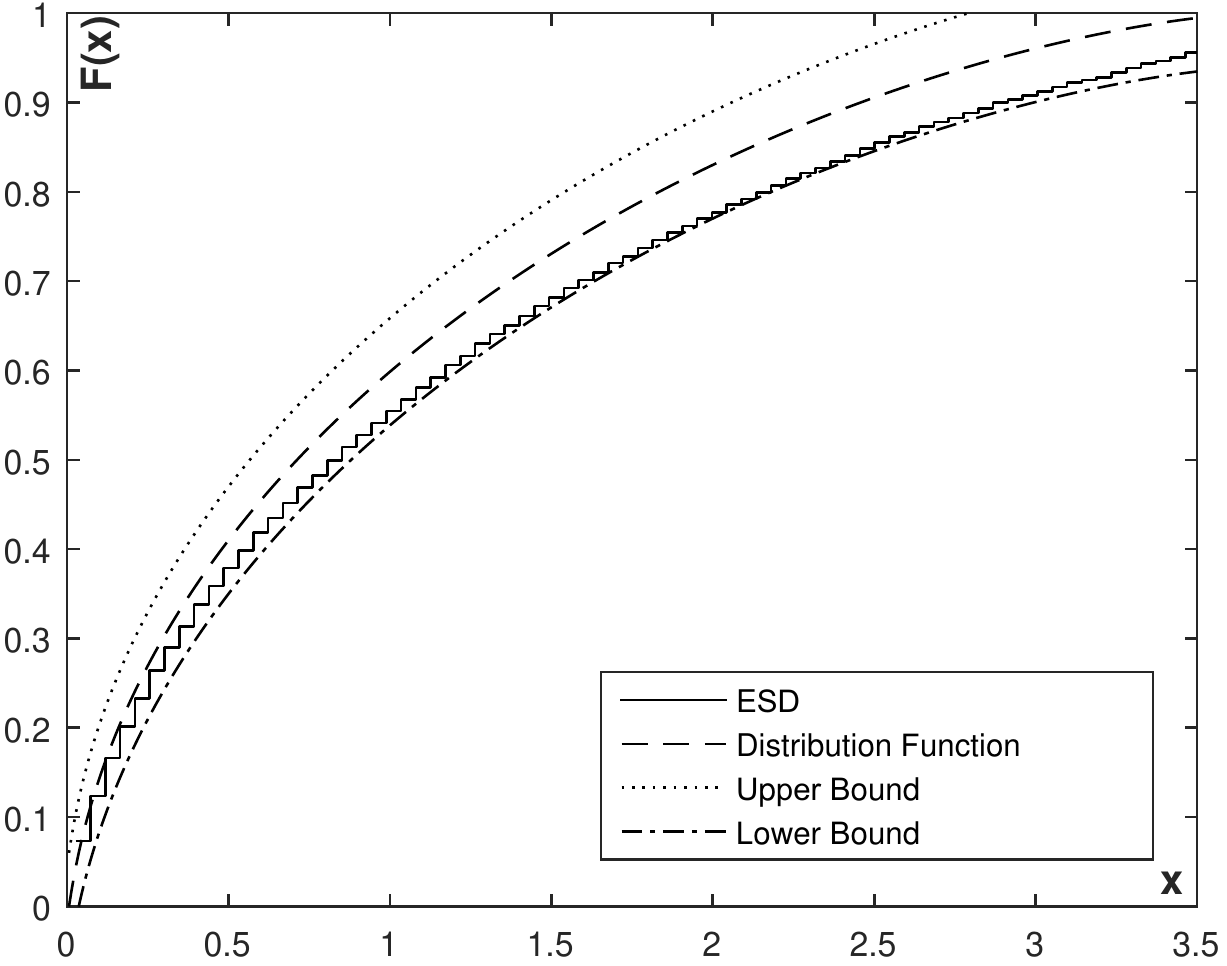}
}
\subfloat[ESD of $\Vector Y_6$ (Abnormal)]{
\includegraphics[width=0.23\textwidth, height=1.3 in]{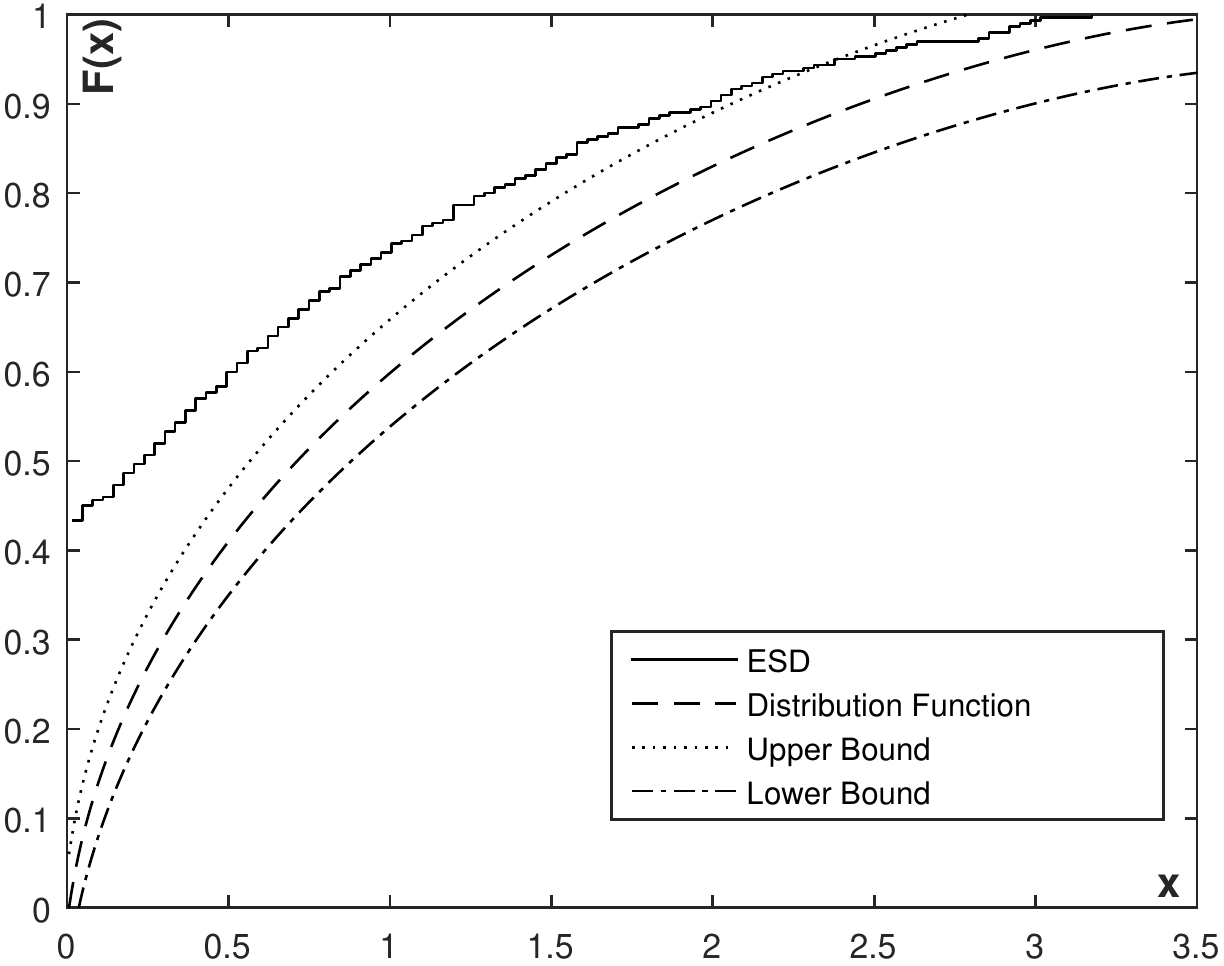}
}
\caption{Anomaly Detection Using LUE matrices}
\label{MP_Law}
\end{figure}

\begin{figure}[H]
\subfloat[Density of $\Vector Z_0$ (Normal)]{
\includegraphics[width=0.23\textwidth, height=1.3 in]{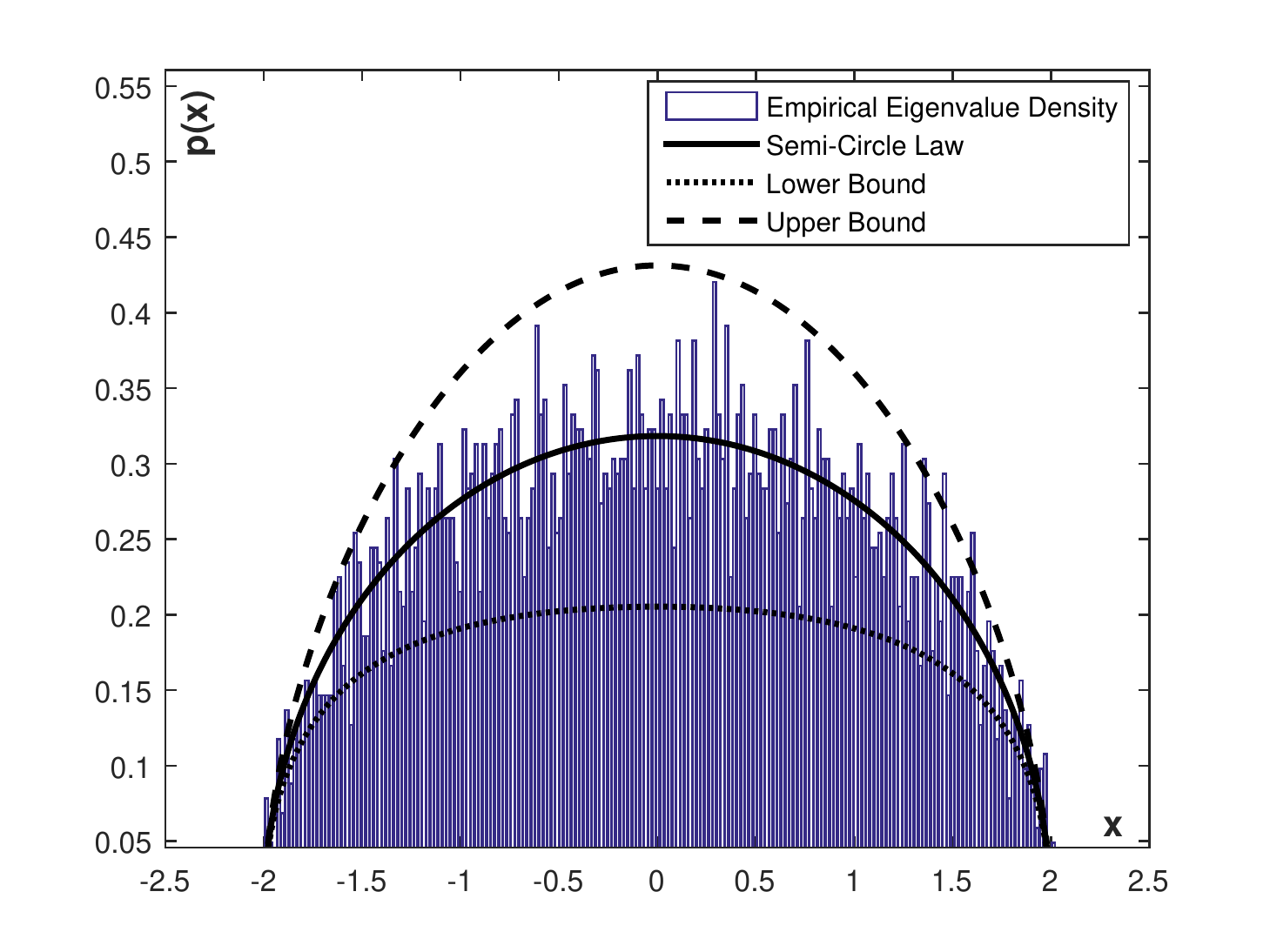}
}
\subfloat[Density of $\Vector Z_6$ (Abnormal)]{
\includegraphics[width=0.23\textwidth, height=1.3 in]{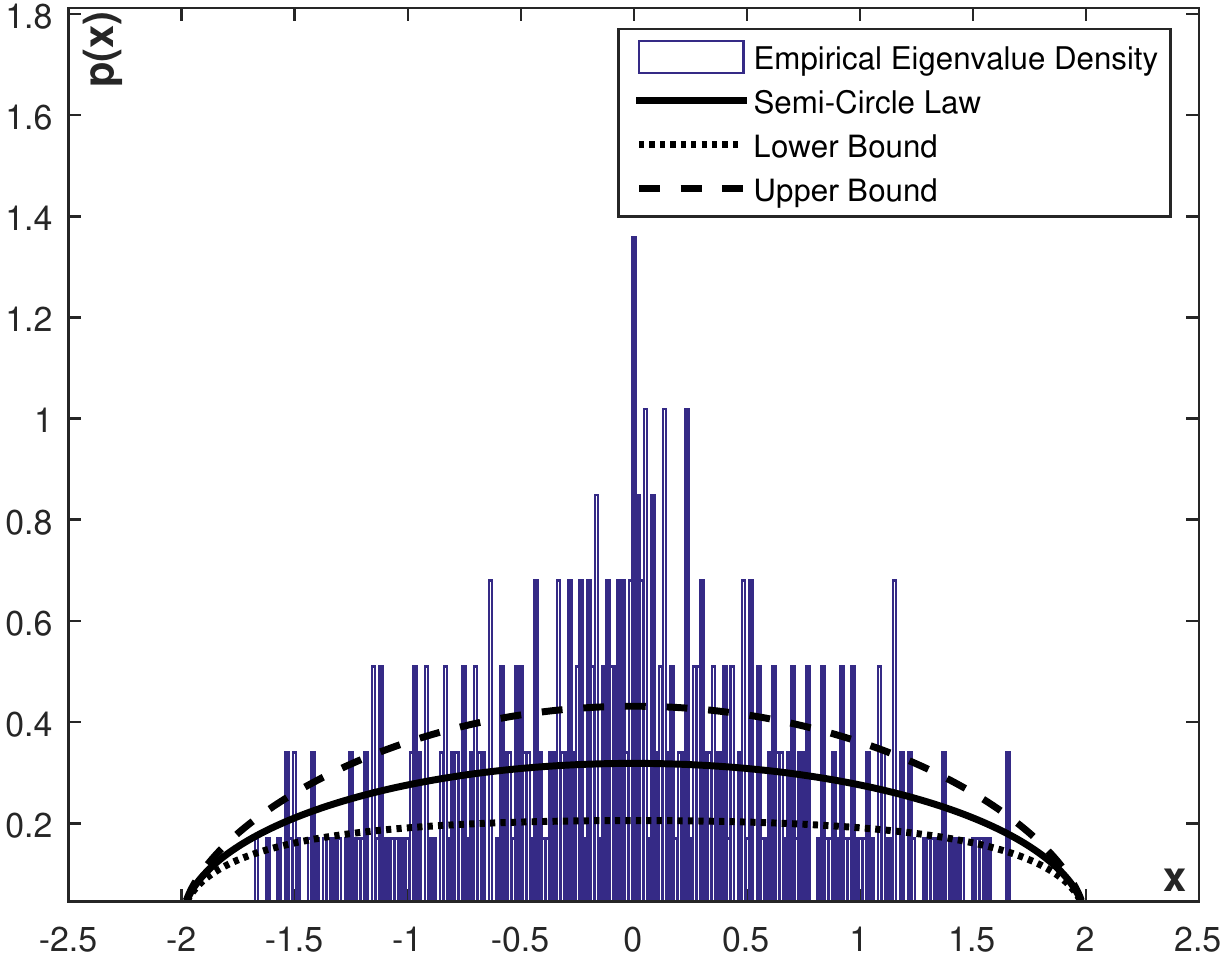}
}

\subfloat[ESD of $\Vector Z_0$ (Normal)]{
\includegraphics[width=0.23\textwidth, height=1.3 in]{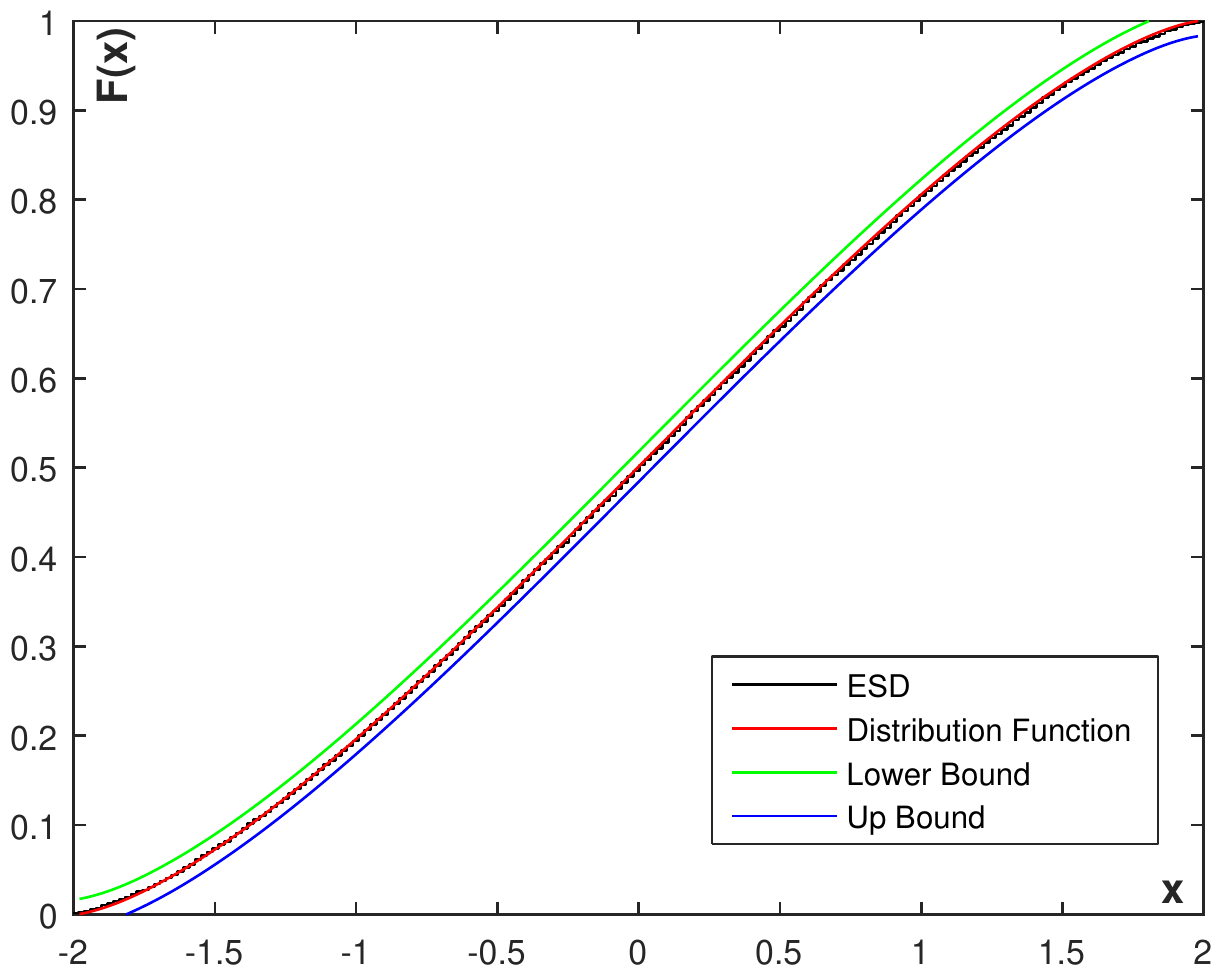}
}
\subfloat[ESD of $\Vector Z_6$ (Abnormal)]{
\includegraphics[width=0.23\textwidth, height=1.3 in]{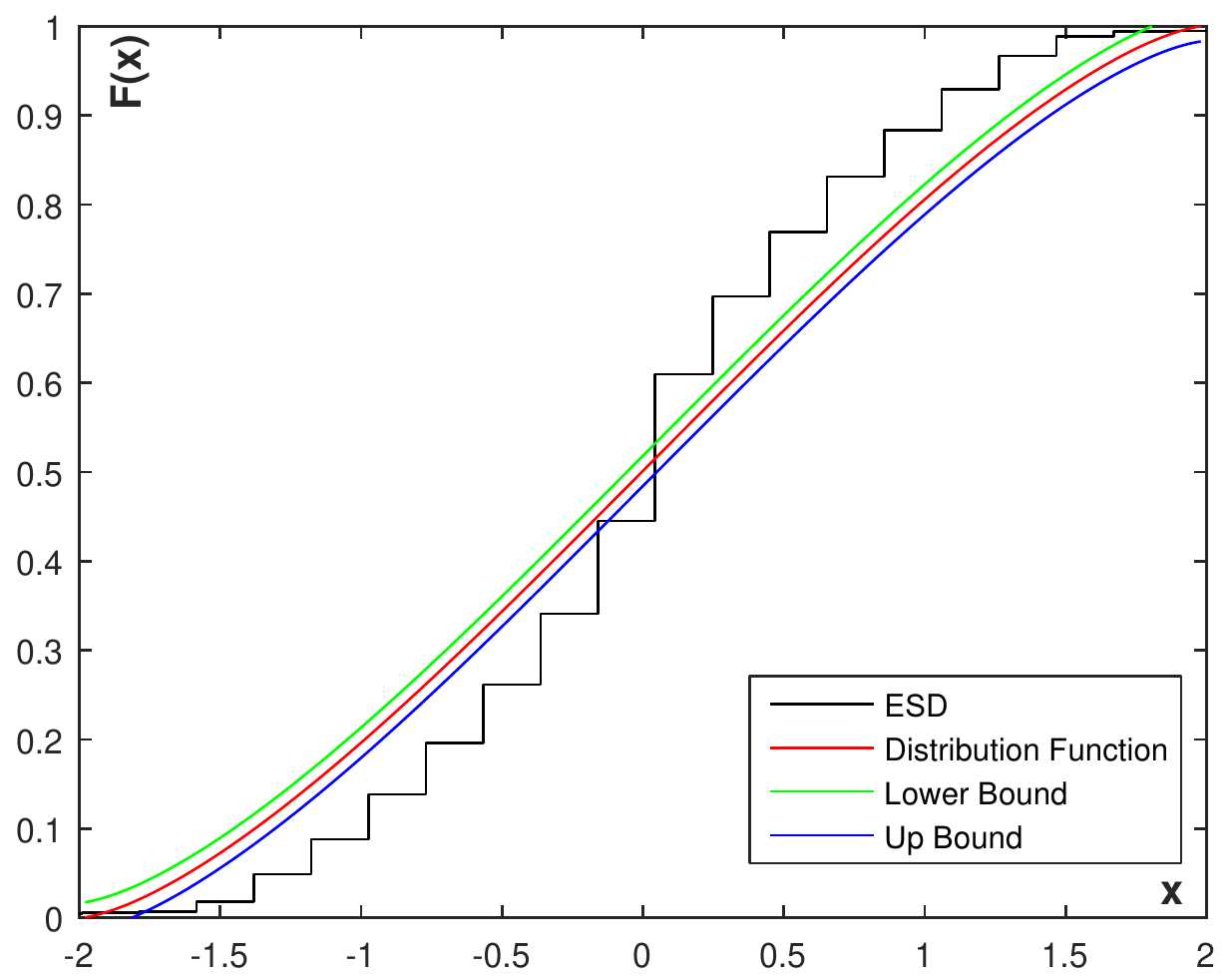}
}
\caption{Anomaly detection using GUE matrices}
\label{Semi_Circle_Law}
\end{figure}

We still set the sampling time at $t_\text{end}\!=\!300$ and $t_\text{end}\!=\!901$.
Following Lemma \ref{thm2:Convergence for Spectra} and Lemma \ref{thm5:Convergence for Spectra},
$\Vector Y_0, \Vector Y_6\Belong{}{\VF R{118}{240}}$ (span $t\!=\![61\!:\!300]$ and $t\!=\![662\!:\!901]$), and
$\Vector Z_0, \Vector Z_6\Belong{}{\VF R{118}{118}}$ (span $t\!=\![183\!:\!300]$ and $t\!=\![784\!:\!901]$)
 are selected. The results are shown in Fig. \ref{MP_Law} and Fig. \ref{Semi_Circle_Law}.
These results validate that empirical spectral density test is competent to conduct anomaly detection---when the power grid is under a normal condition, the empirical spectral density ${f_{\bf{A}}}\left( x \right)$ and the ESD function ${F_{\bf{A}}}\left( x \right)$ are almost strictly bounded between the upper bound and the lower bound of their asymptotic limits. On the other hand, these results also  validate that GUE and LUE are proper mathematical tools to model the power grid operation.


\subsection{Steady Stability Evaluation}
\label{sec:case0stabilityanalysis}
\begin{figure}[htbp]
\centering
\subfloat[$V-P$ Curve]{\label{fig:Case0NosecurveVP}
\includegraphics[width=0.23\textwidth]{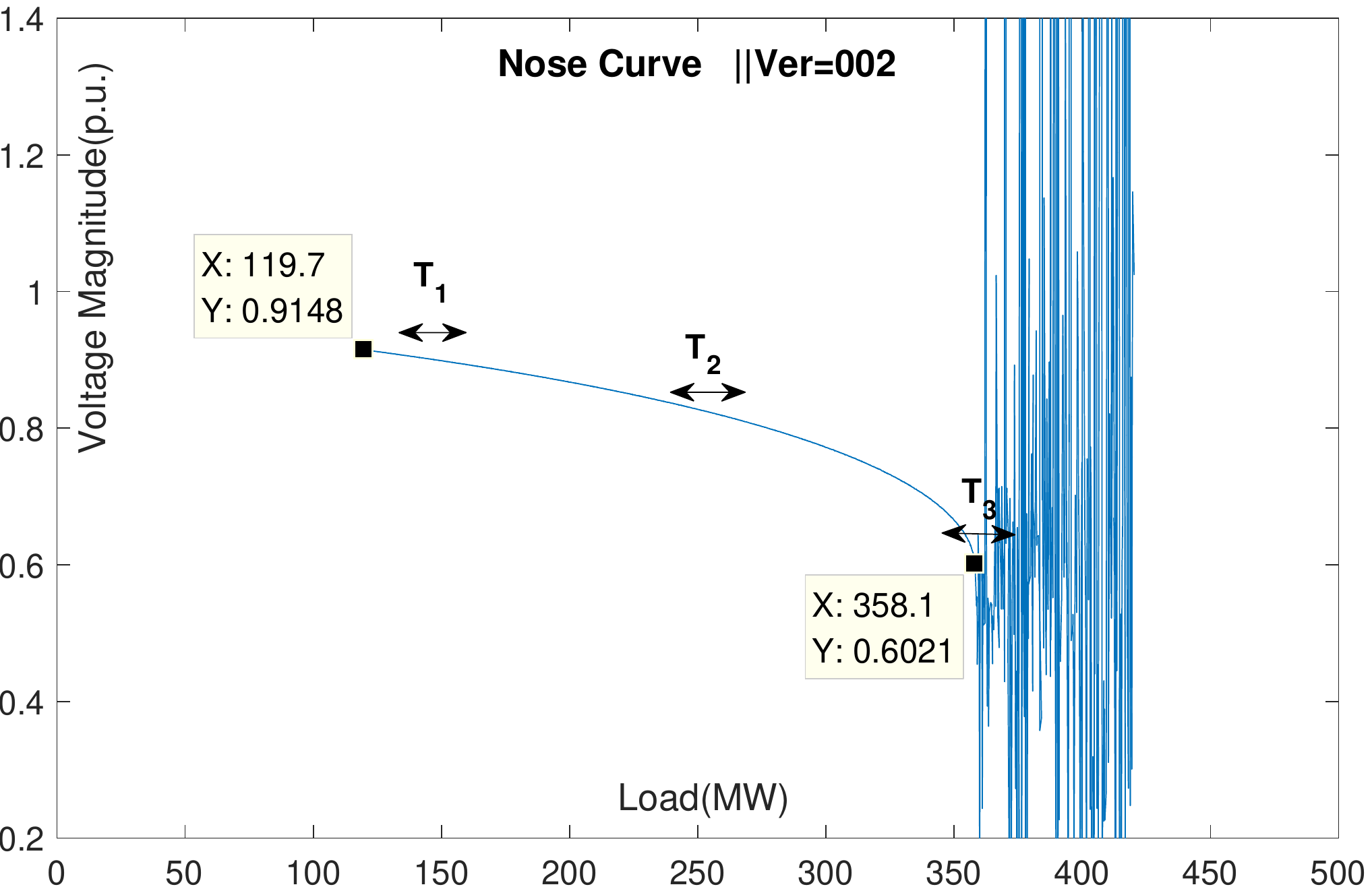}
}
\subfloat[$\lambda-P$ Curve]{\label{fig:Case0NosecurveEigP}
\includegraphics[width=0.23\textwidth]{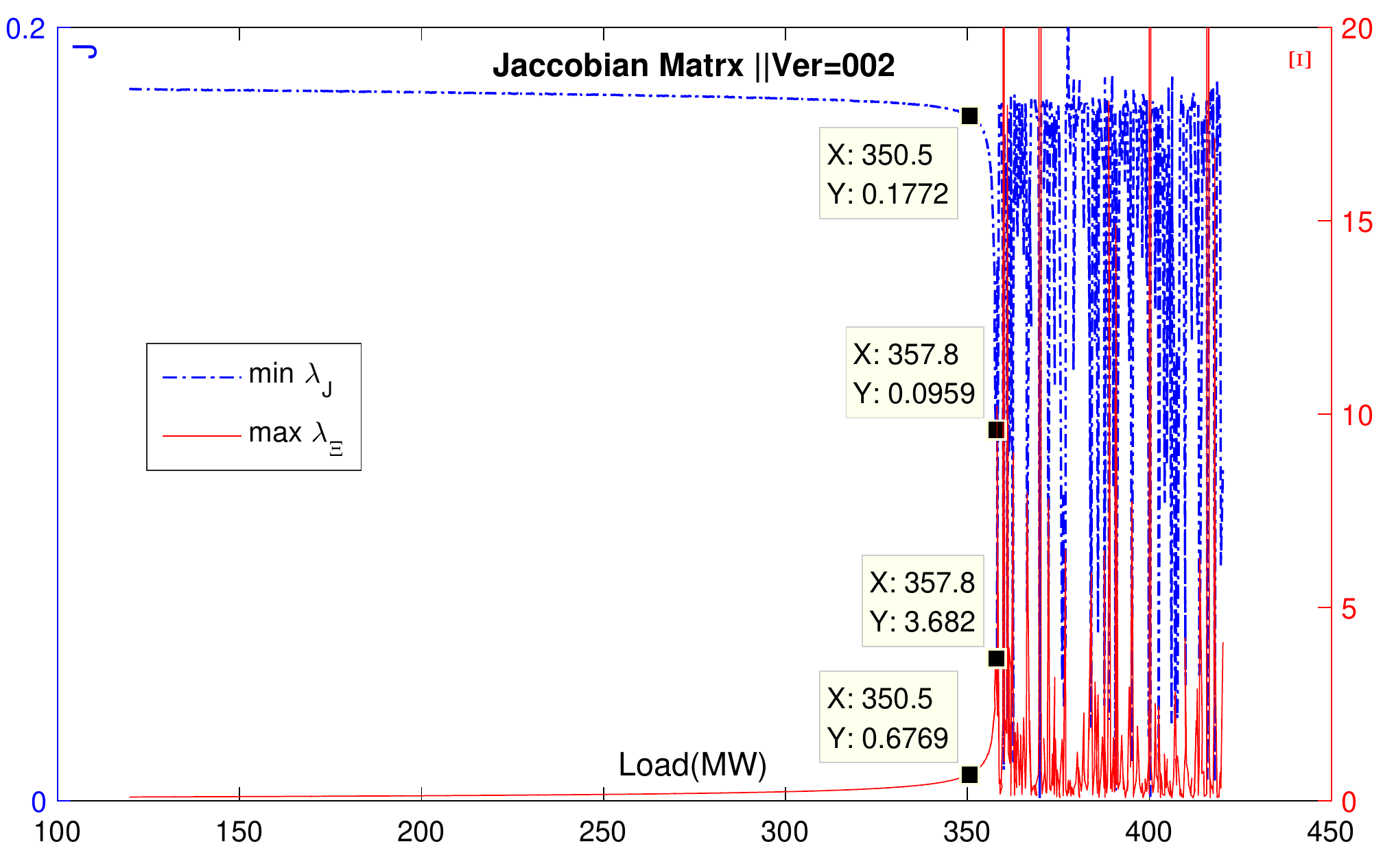}
}
\caption{The $V-P$ curve and  $\lambda-P$ curve.}
\label{fig:Case0C}
\end{figure}

\begin{figure}[htbp]
\centering
\subfloat[Ring Law for $\Vector T_1$]{\label{fig:Case0T1ring}
\includegraphics[width=0.20\textwidth]{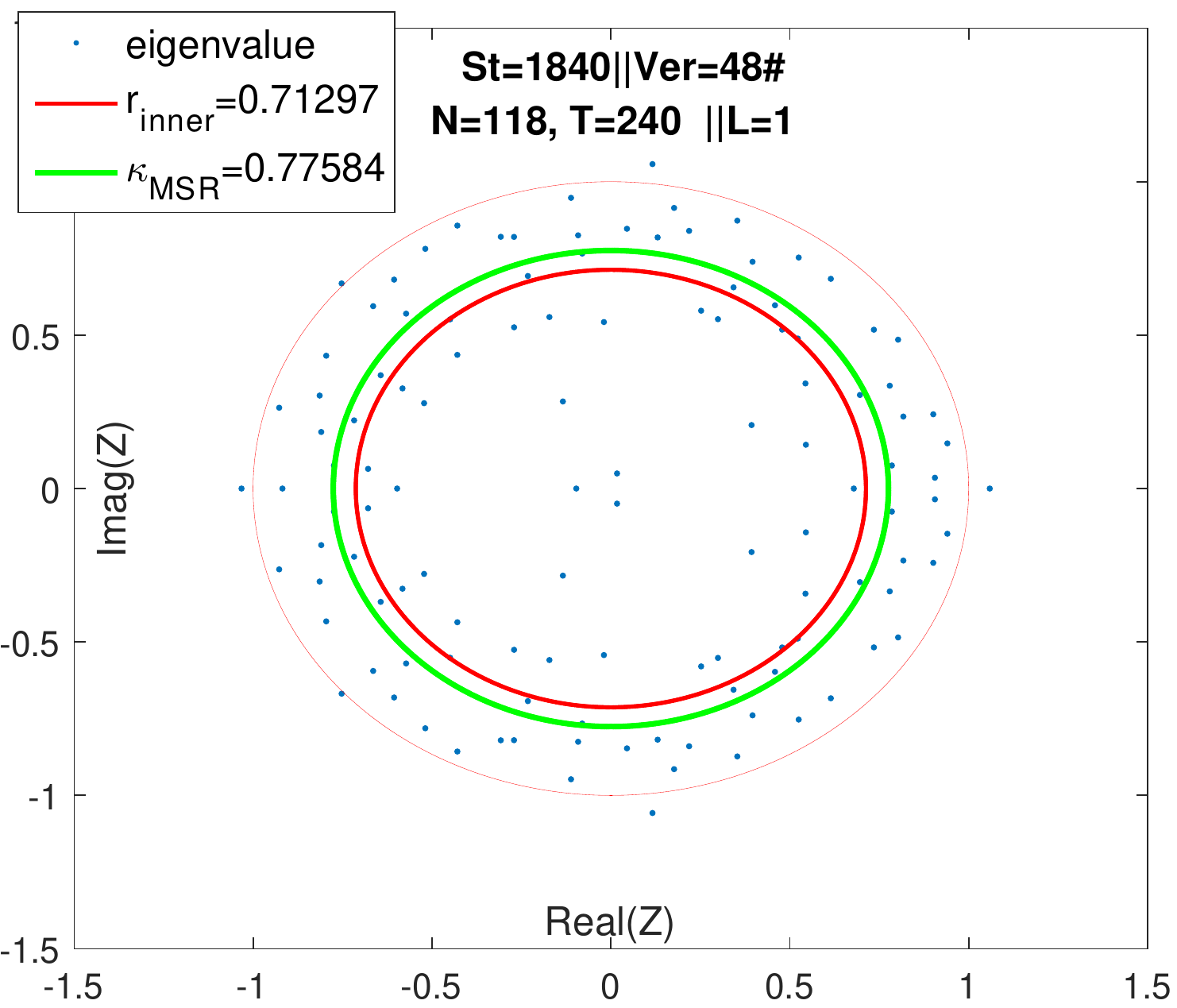}
}
\subfloat[M-P Law for $\Vector T_1$]{\label{fig:Case0T1mp}
\includegraphics[width=0.26\textwidth]{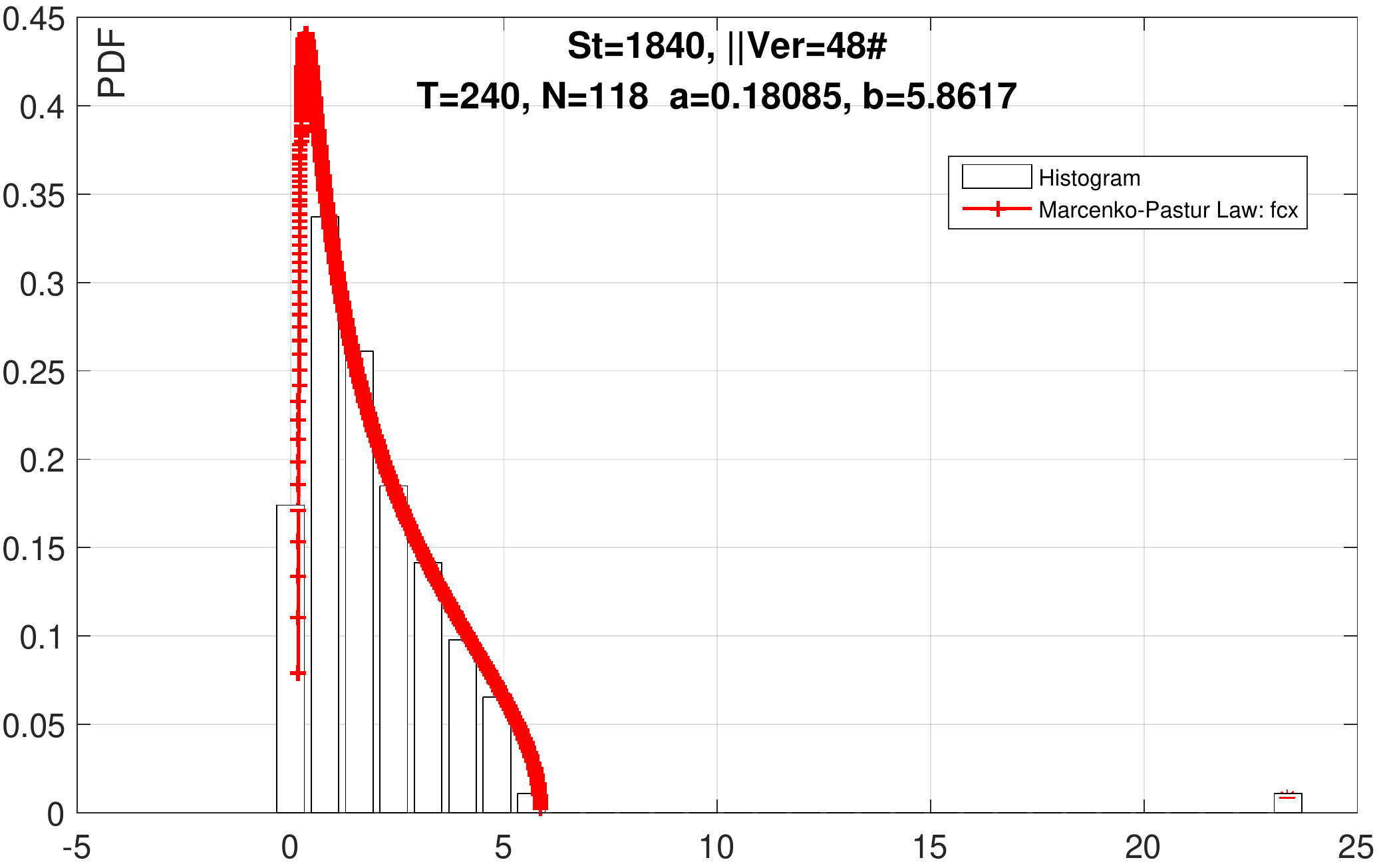}
}

\subfloat[Ring Law for $\Vector T_2$]{\label{fig:Case0T2ring}
\includegraphics[width=0.20\textwidth]{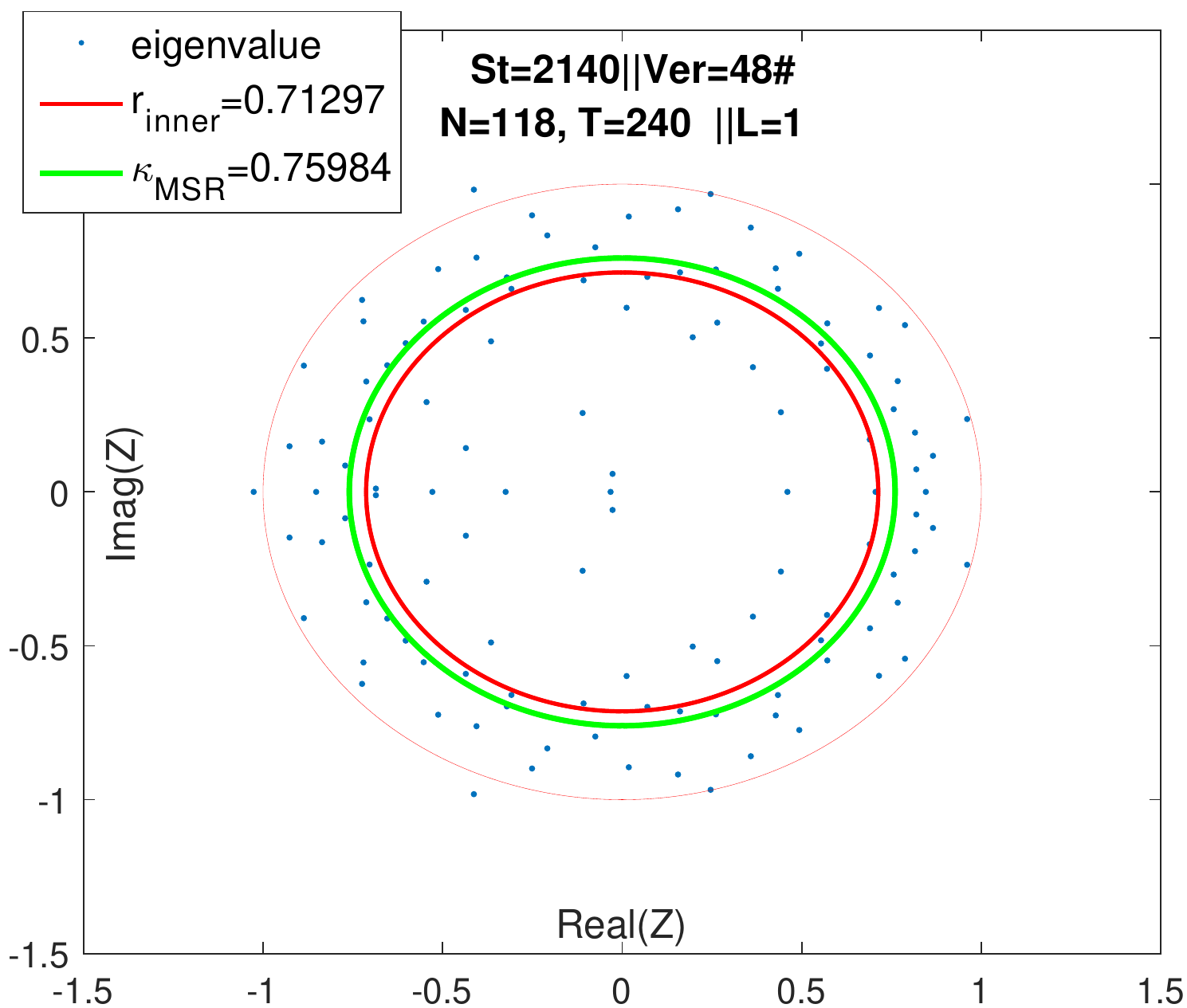}
}
\subfloat[M-P Law for $\Vector T_2$]{\label{fig:Case0T2mp}
\includegraphics[width=0.26\textwidth]{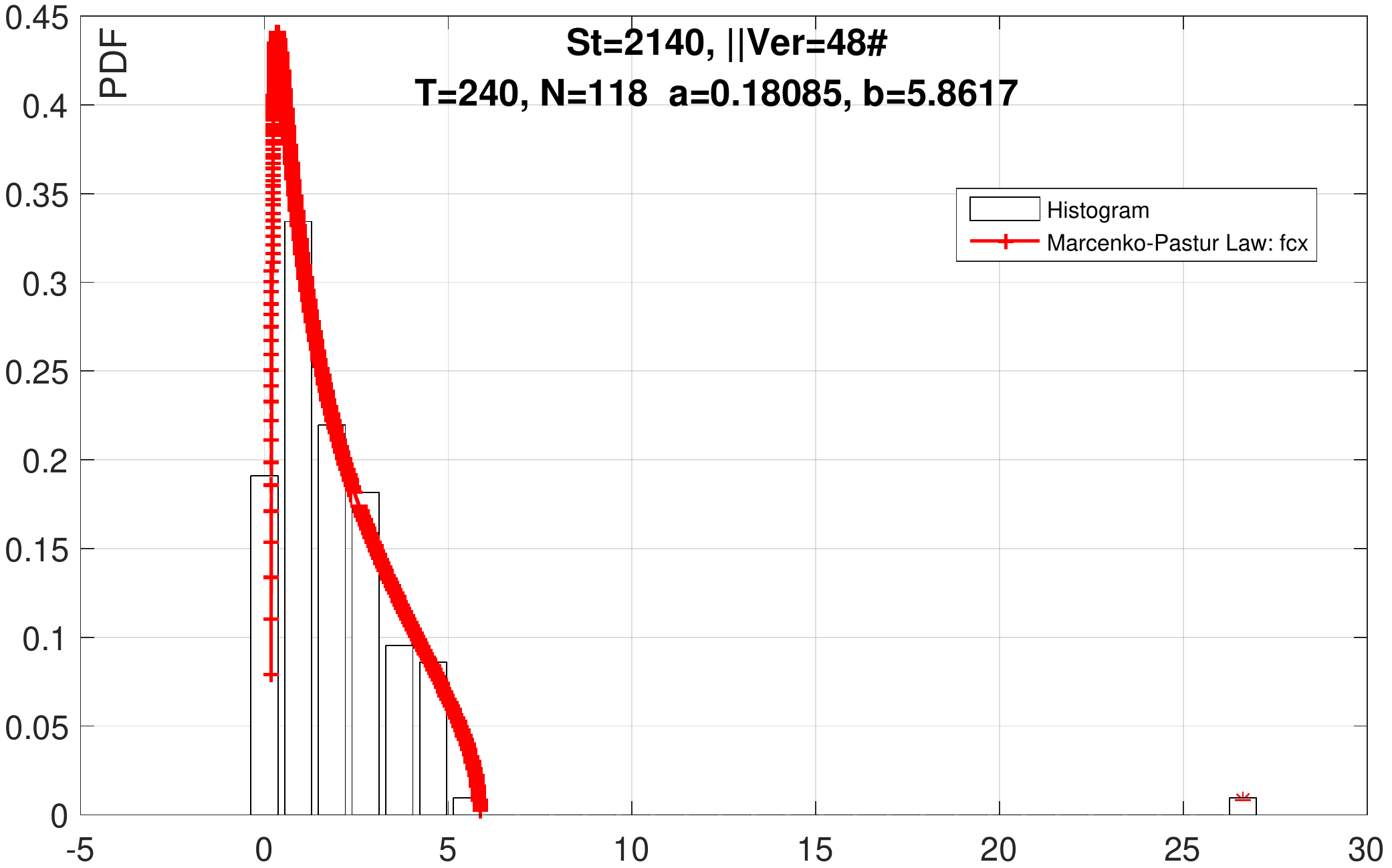}
}

\subfloat[Ring Law for $\Vector T_3$]{\label{fig:Case0T3ring}
\includegraphics[width=0.20\textwidth]{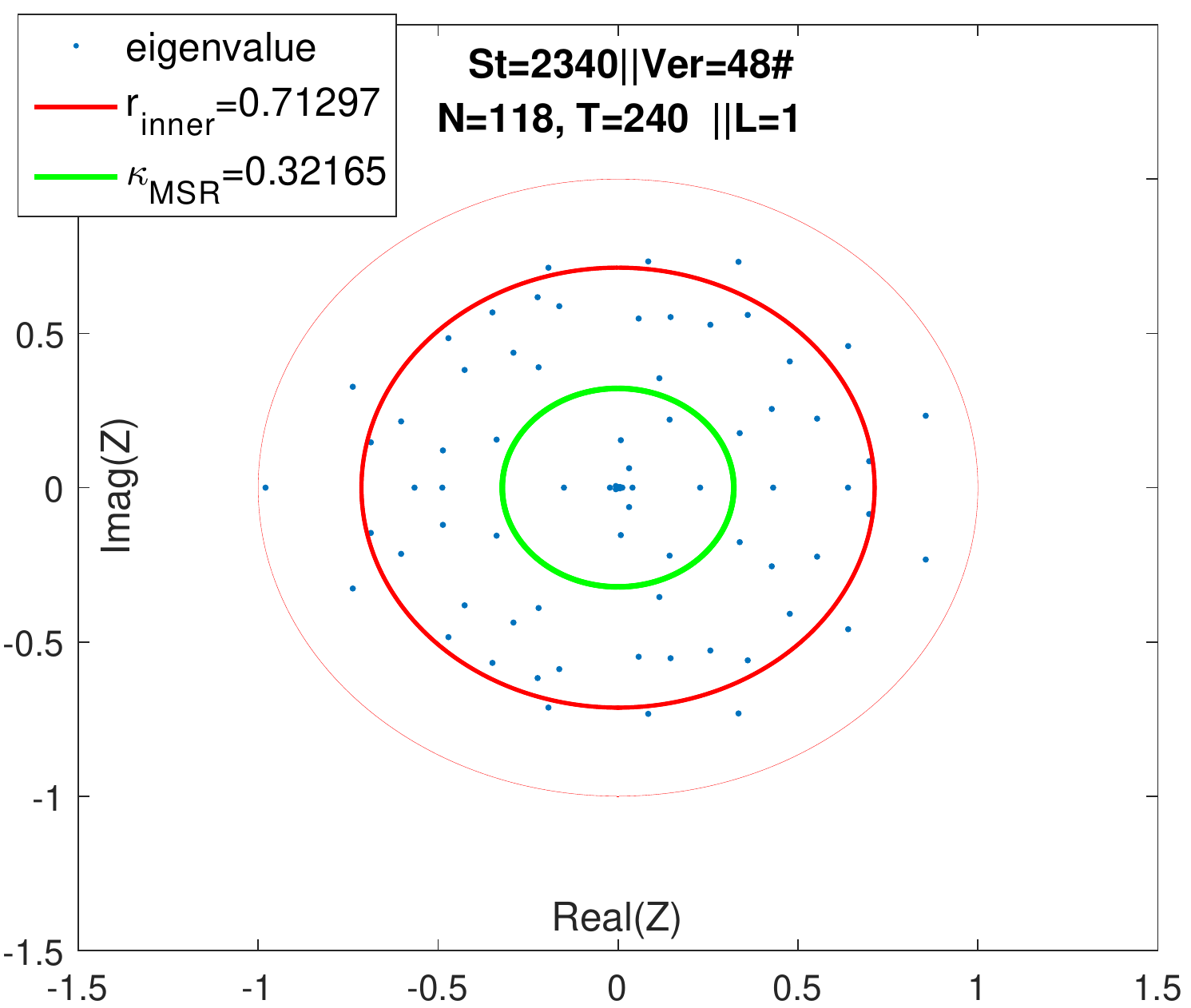}
}
\subfloat[M-P Law  for $\Vector T_3$]{\label{fig:Case0T3mp}
\includegraphics[width=0.26\textwidth]{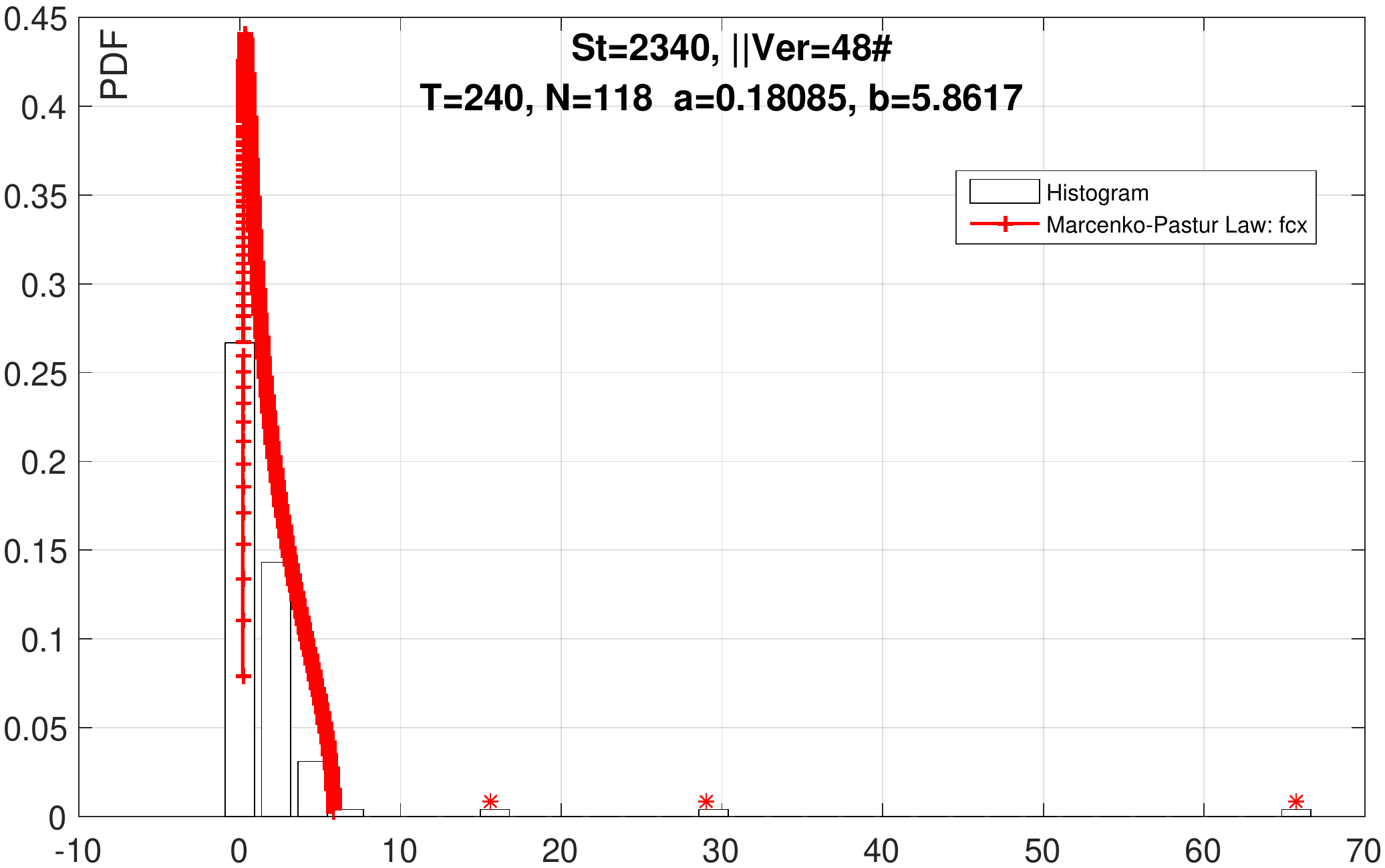}
}

\caption{RMT-based results for voltage stability evaluation.}
\label{fig:Case0F}
\end{figure}

The $V-P$ curve (also called nose curve) and the smallest eigenvalue of the Jacobian Matrix \cite{lim2016svd} are two clues for steady stability evaluation.   In this case, we focus on  \textbf{E}4 stage during which \VPbus{52} keeps increasing until the system exceeds its steady stability limit. The $V-P$ curve and  $\lambda-P$ curve, respectively, are given in Fig. \ref{fig:Case0NosecurveVP} and Fig. \ref{fig:Case0NosecurveEigP}. Furthermore, we choose some data section, $\mathbf T_1\!:\![1601\!:\!1840]; \quad \mathbf T_2\!:\![1901\!:\!2140]; \quad \mathbf T_3\!:\![2101\!:\!2340],$ shown as Fig. \ref{fig:Case0NosecurveVP}. The RMT-based results are shown as Fig. \ref{fig:Case0F}. The outliers become more evident as the stability degree decreases. The statistics of the outliers,  in some sense, are similar to the smallest eigenvalue of Jacobian Matrix, Lyapunov Exponent or the entropy.

For further analysis, we take the signal and stage division into account. In general, sorted by the stability degree, the stages are ordered as  $\textbf{S}0>\textbf{S}1>\textbf{S}2 \gg \text{max}(\textbf{S}3, \textbf{S}4, \textbf{S}5)>\text{min}(\textbf{S}3, \textbf{S}4, \textbf{S}5)\gg \textbf{S}6 \gg \textbf{S}7$. According to Fig. \ref{fig:Case0LESs}, we make the Table \ref{Tab: Case0LESs}. The high-dimensional indicators $\overline{\tau{}_\mathbf{X}}_\text{R}$ has the same trend as the stability degree order. These statistics have the potential for data-driven stability evaluation. Moreover, based on the Gaussian property of LES indicators, hypothesis tests are designed for the anomaly detection; see \cite{he2017invisable} for details.

\begin{table}[htbp]
\caption{Indicator of Various LESs at Each Stage.}
\label{Tab: Case0LESs}
\centering

\begin{minipage}[!h]{0.49\textwidth}
\centering

\footnotesize
\begin{tabularx}{\textwidth} { >{\scshape}l !{\color{black}\vrule width1pt}  >{$}l<{$}  >{$}l<{$}  >{$}l<{$}   >{$}l<{$}  >{$}l<{$}  >{$}l<{$}   >{$}l<{$}  >{$}l <{$} }   
\toprule[1.5pt]
\hline
 & {\text{MSR}} &   {\text{T}_2} & {\text{T}_3} & {\text{T}_4} & {\text{DET}}  & {\text{LRF}} & \\
\hline
\hline
\multicolumn{7}{l} {$\textbf{E}_0$: Theoretical Value}\\
\hline
\STE{\tau}&0.8645&1338.3&10069 &8.35\text{E}4&48.322&73.678 \\
\toprule[1pt]

\multicolumn{7}{l} {$\textbf{S}0$ [{0240:0500}, {261}]: Small fluctuations around 0 MW} \textcircled{1}\\
\hline
$\overline{\tau{}_\mathbf{X}}_\text{R}$&0.995&1.010&1.040&1.080&0.959&1.014\\
\toprule[1pt]

\multicolumn{7}{l} {$\textbf{S}5$ [{0501:0739}, {239}]: A step signal (0 MW $\uparrow$ 30 MW) is included} \textcircled{4}\\
\hline
$\overline{\tau{}_\mathbf{X}}_\text{R}$&0.9331&1.280&2.565&7.661&0.5453&1.284\\
\toprule[1pt]

\multicolumn{7}{l} {$\textbf{S}1$ [{0740:0900}, {161}]: Small fluctuations around 30 MW} \textcircled{2}\\
\hline
$\overline{\tau{}_\mathbf{X}}_\text{R}$&0.9943&1.010&1.039&1.084&0.9568&1.015\\
\toprule[1pt]

\multicolumn{7}{l} {$\textbf{S}6$ [{0901:1139}, {239}]: A step signal (30 MW $\uparrow$ 120 MW) is included} \textcircled{7}\\
\hline
$\overline{\tau{}_\mathbf{X}}_\text{R}$&0.8742&2.054&1.06\text{E}1&7.22\text{E}1&7\text{E}\!-\!2&1.597\\
\toprule[1pt]
\multicolumn{7}{l} {$\textbf{S}2$ [{1140:1300}, {161}]: Small fluctuations around 120 MW} \textcircled{3}\\
\hline
$\overline{\tau{}_\mathbf{X}}_\text{R}$&0.9930&1.019&1.067&1.135&0.9488&1.021\\
\toprule[1pt]
\multicolumn{7}{l} {$\textbf{S}4$ [{1301:1539}, {239}]: A ramp signal (119.7 MW $\nearrow$) is included} \textcircled{4}\\
\hline
$\overline{\tau{}_\mathbf{X}}_\text{R}$&0.9337&1.295&2.787&9.615&0.5316&1.294\\
\toprule[1pt]
\multicolumn{7}{l} {$\textbf{S}3$ [{1540:2253}, {714}]: Steady increase ($\nearrow$ 358.1 MW)} \textcircled{4}\\
\hline
$\overline{\tau{}_\mathbf{X}}_\text{R}$&0.8906&1.717&6.530&3.48\text{E}1&0.1483&1.545\\
\toprule[1pt]
\multicolumn{7}{l} {$\textbf{S}7$ [{2254:2500}, {247}]: Static voltage collapse (361.9 MW $\nearrow$)} \textcircled{8}\\
\hline
$\overline{\tau{}_\mathbf{X}}_\text{R}$&0.4259&1.02\text{E}1&2.11\text{E}2&4.65\text{E}3&-1.4\text{E}1&1.08\text{E}1\\
\toprule[1pt]

\hline
\toprule[1pt]
\end{tabularx}
\raggedright
 {\small{}
 *$\overline{\tau{}_\mathbf{X}}_\text{R}=\overline{\tau{}_\mathbf{X}}/\mathbb{E}({\tau})$.
 \normalsize{}
}
\end{minipage}
\end{table}

\subsection{Correlation Analysis}
\label{sec:case0sensitivefactor}
\begin{figure}[htbp]
\centering
\includegraphics[width=0.45\textwidth]{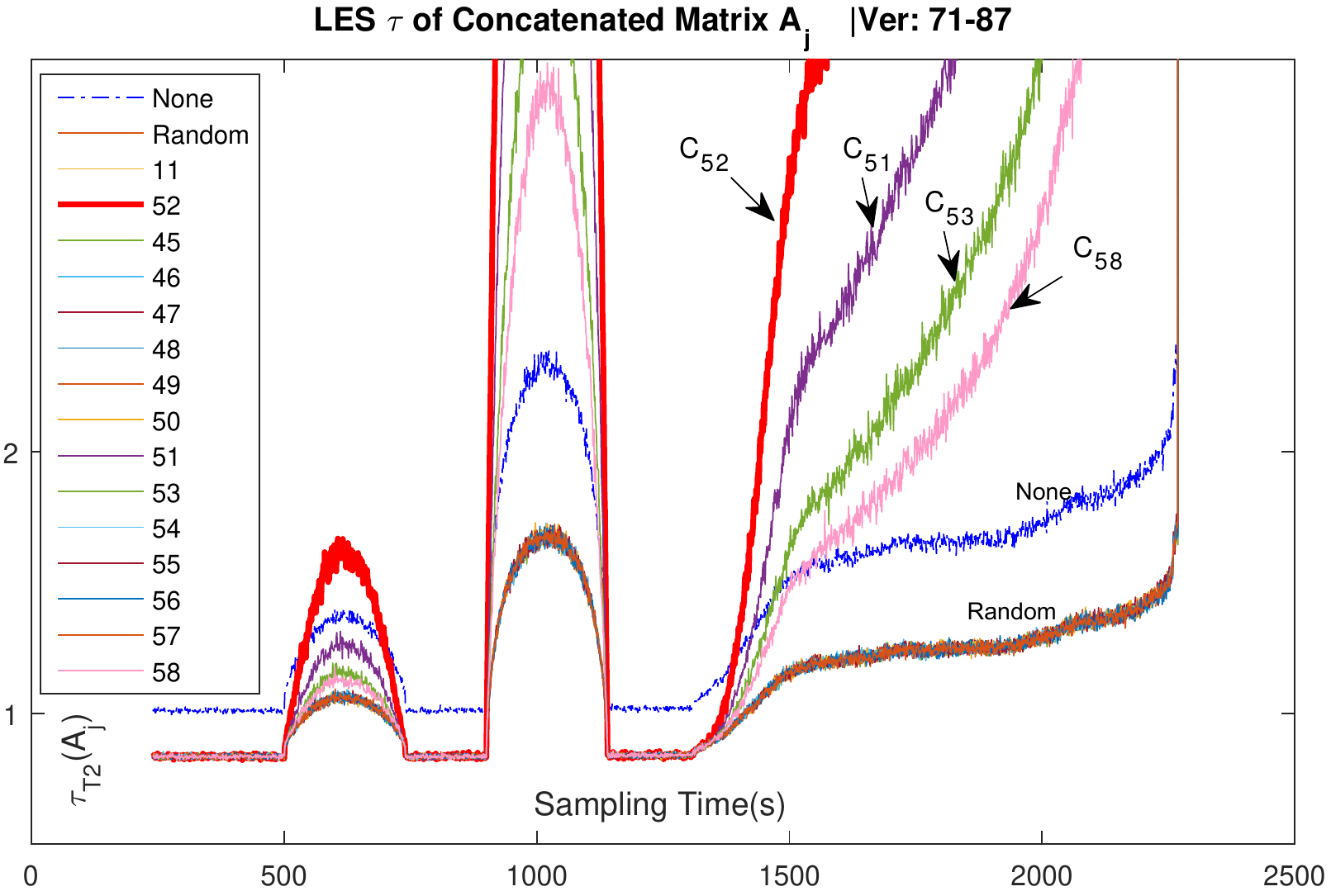}
\caption{Sensitivity Analysis based on Concatenated Matrix.}
\label{fig:Case0Concatenation}
\end{figure}

The key for correlation analysis is the  concatenated matrix $\mathbf{A}_i$, which consist of two part---the basic matrix $\mathbf{B}$ and a certain factor matrix $\mathbf{C}_i$, i.e., $\mathbf{A}_i\!=\![\mathbf{B}; \mathbf{C}_i]$. For details, see our previous work \cite{he2015corr}.
The LES of each $\mathbf{A}_i$ is computed in parallel, and Fig. \ref{fig:Case0Concatenation} shows the results.

In Fig. \ref{fig:Case0Concatenation}, the blue dot line (marked with None) shows the LES of basic matrix  $\mathbf{B}$, and the orange line (marked with Random) shows the LES of the concatenated matrix $[\mathbf{B}; \mathbf{R}]$ ($\mathbf{R}$ is the standard Gaussian Random Matrix). Fig. \ref{fig:Case0Concatenation} demonstrates that: 1) node 52 is the causing factor of the anomaly; 2) sensitive nodes are 51, 53, and 58; and 3) nodes 11, 45, 46, etc, are not affected by the anomaly. Based on this algorithm, we can conduct behavior analysis, e.g.,  detection and estimation of residential PV installations \cite{7456317}. It is another hot topic, and we expand it in \cite{he2017invisable}.

\subsection{SA with Asynchronous Data}
\label{sec:case0advantageous}
The proposed data-driven method is robust against bad data both in space and in time. In our previous work \cite{he2015les}, we have successfully conducted SA with data loss in the core area. This part we talk about SA with asynchronous data. It is common that asynchronous data exists in the data platforms such as SCADA or WAMS. The problem is mainly caused by erroneous time-tags or communication delays. Sometimes, for a certain signal, the proper delay protection or interaction/response mechanism will also lead to asynchronous data.  It is hard to measure or even detect the time delay via traditional methods. Our approach has a special meaning here.

\begin{figure}[htbp]
\centering
\includegraphics[width=.38\textheight]{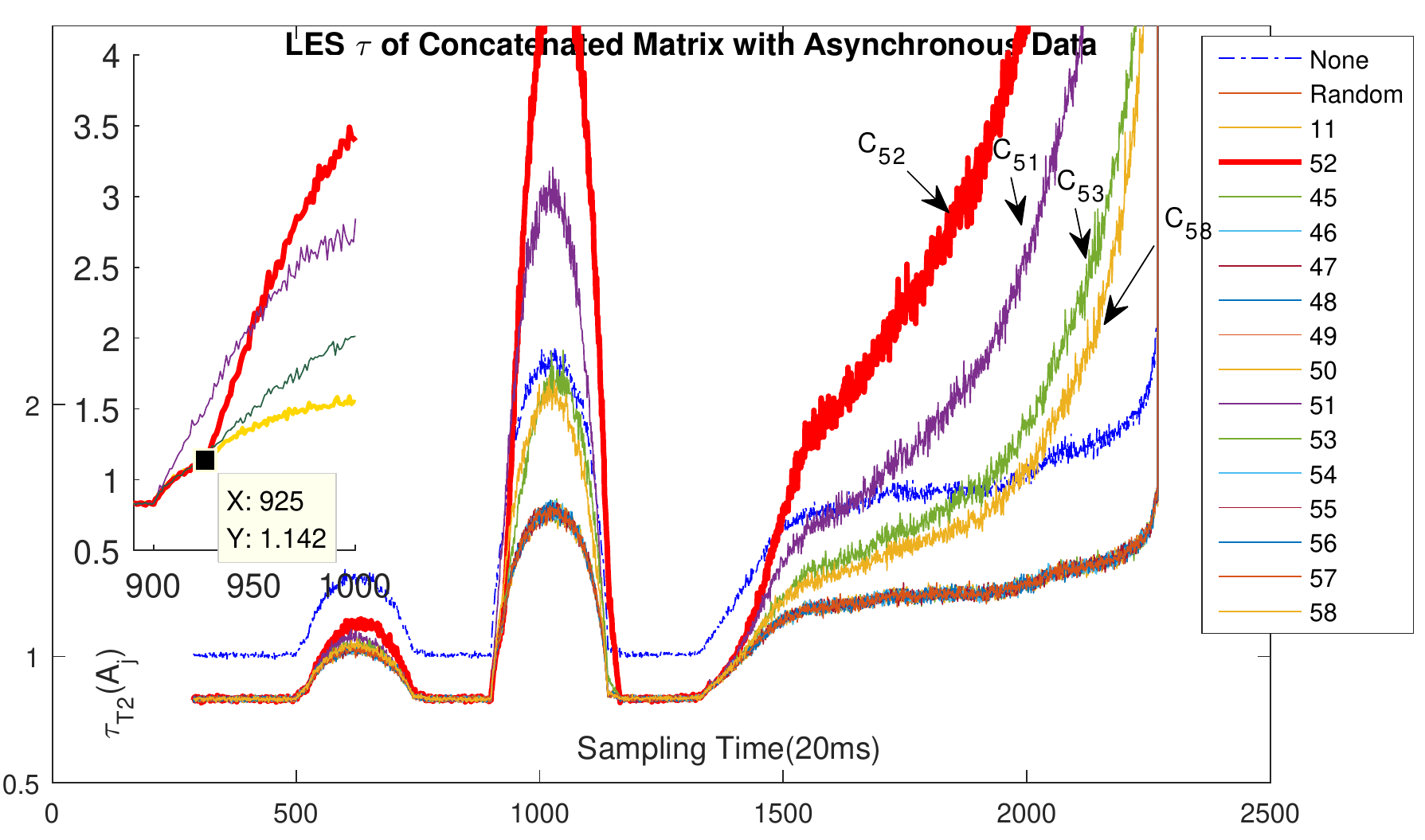}
\caption{Situation awareness with asynchronous data.}
\label{fig:SAunsyn}

\end{figure}

Using the simulated data, we make an artificial delay of 25 sampling points for 7 nodes---11, 14, 50, 52, 53, 77, and 81. With the concatenation operation introduced above, similarly, we obtain the results shown as Fig. \ref{fig:SAunsyn}. It is an interesting discovery that the approach is robust against asynchronous data: 1) the anomalies are detected at $t\!=\!501$ and $t\!=\!901$; 2) node 52 is the most sensitive node; 3) with more detailed observation, we can even quantitatively draw the conclusion that there exists a 25 sampling points delay ($925-900$) for node 52.  It is surprising that the exact delay value can be recovered for the particular node! The power of our proposed approach is vividly exhibited here.

\section{Case Studies Using Field Data}
\label{section: case2}

\begin{figure}[htbp]
\centering
\includegraphics[width=0.47\textwidth]{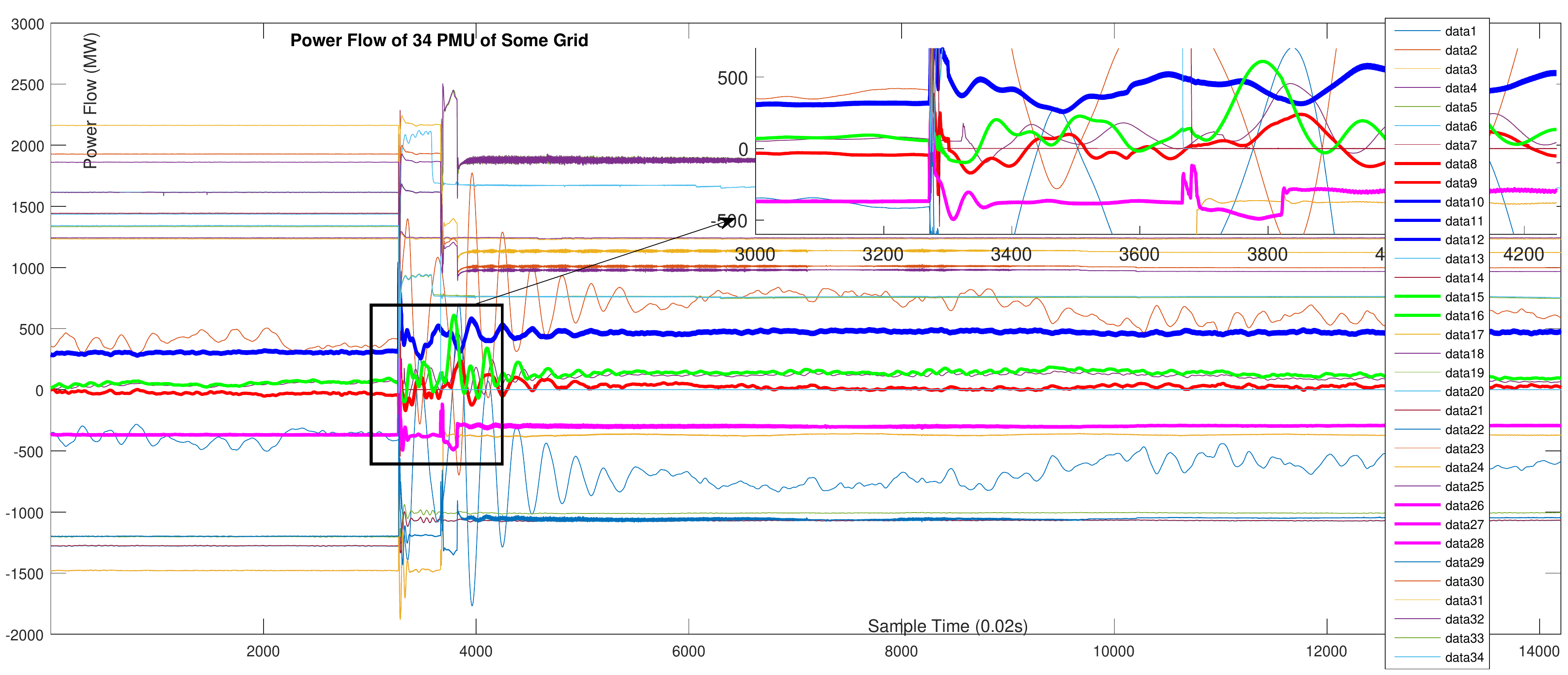}
\caption{Raw power flow data of 34 PMUs.}
\label{Fig:Case2Grid}
\end{figure}

\begin{figure}[H]
\centering
\subfloat[Pre-fault: Ring Law]{\label{fig:Case0T1ring}
\includegraphics[width=0.22\textwidth]{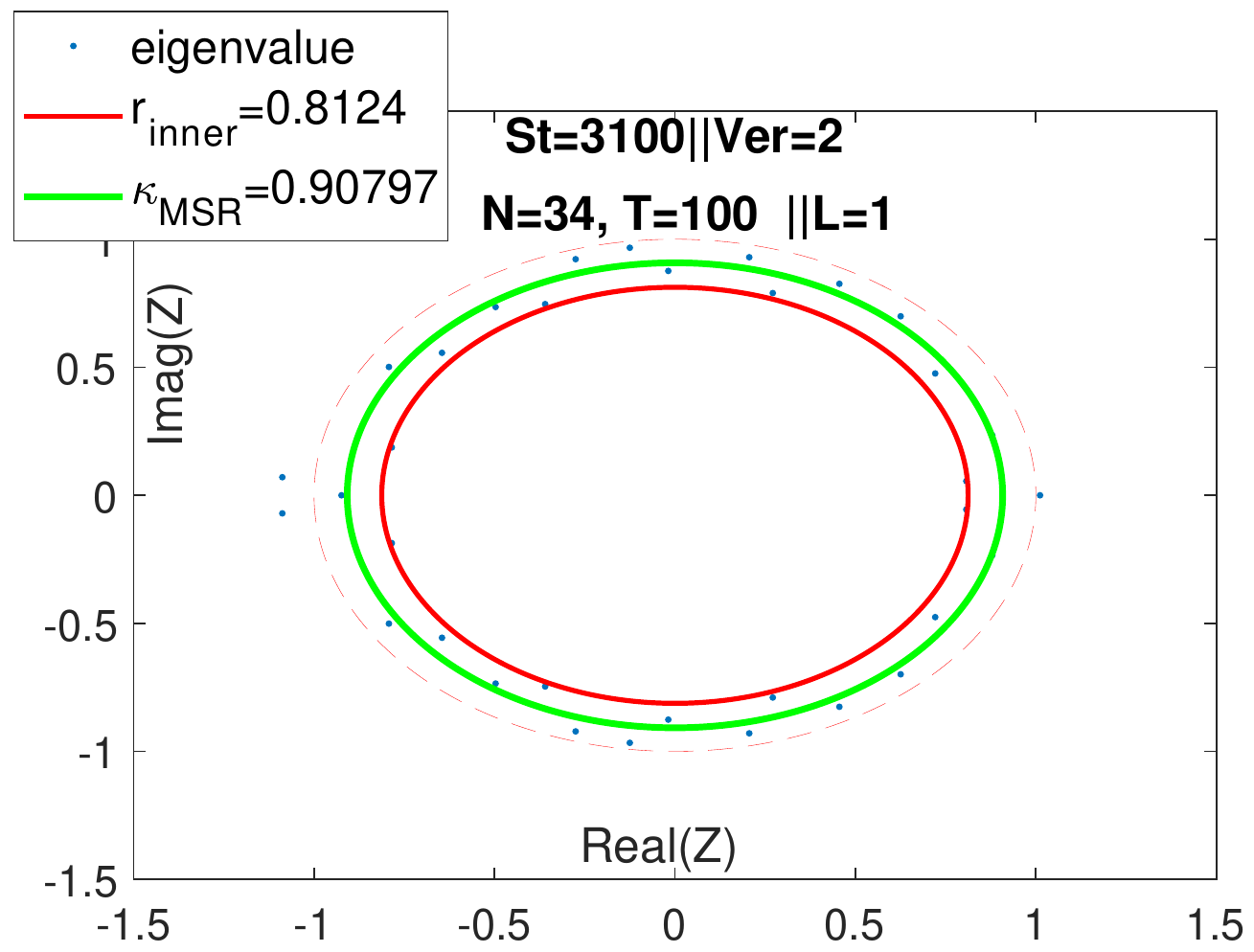}
}
\subfloat[Pre-fault: M-P Law]{\label{fig:Case0T1mp}
\includegraphics[width=0.20\textwidth]{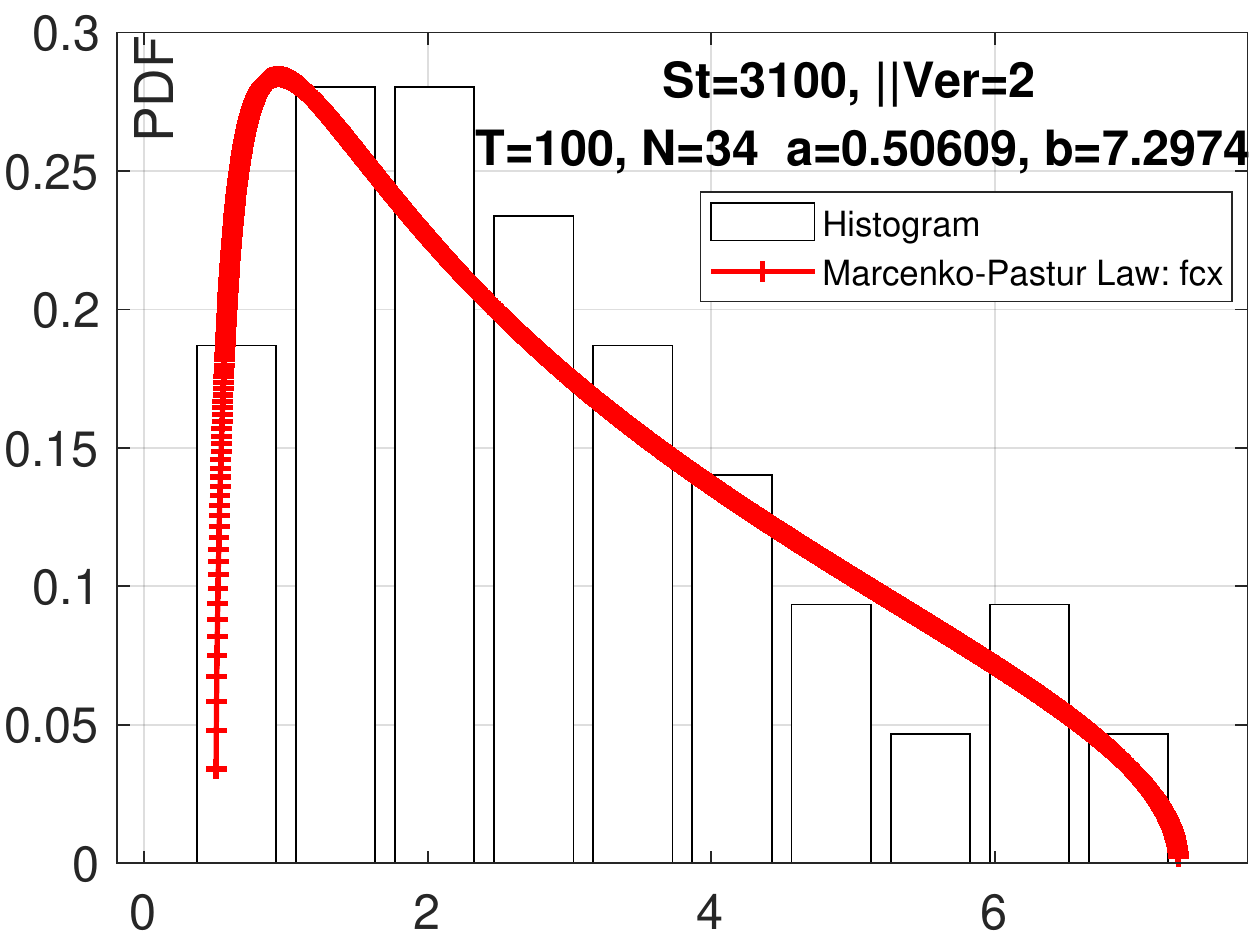}
}

\subfloat[During fault: Ring Law]{\label{fig:Case0T2ring}
\includegraphics[width=0.22\textwidth]{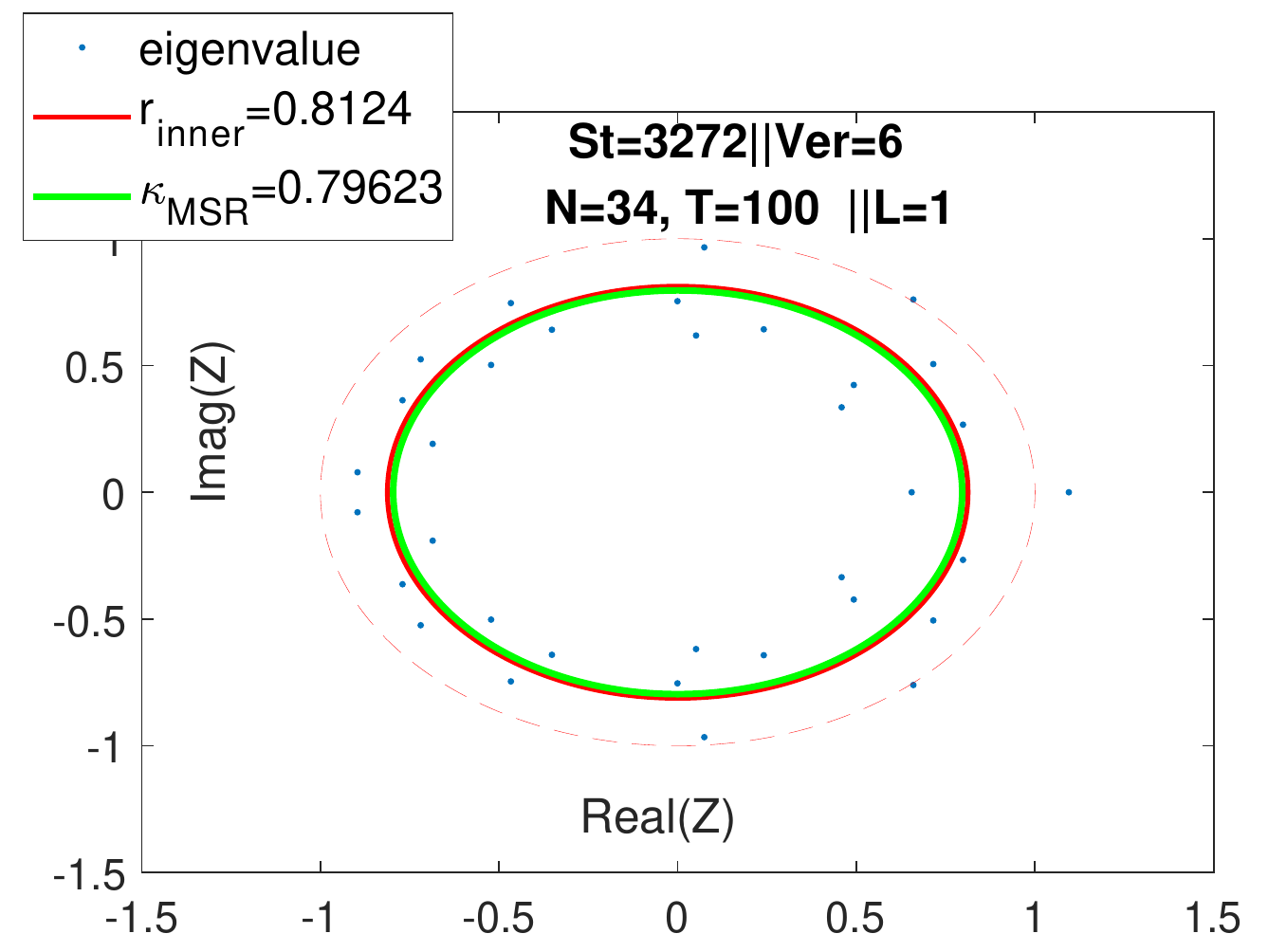}
}
\subfloat[During fault: M-P Law]{\label{fig:Case0T2mp}
\includegraphics[width=0.20\textwidth]{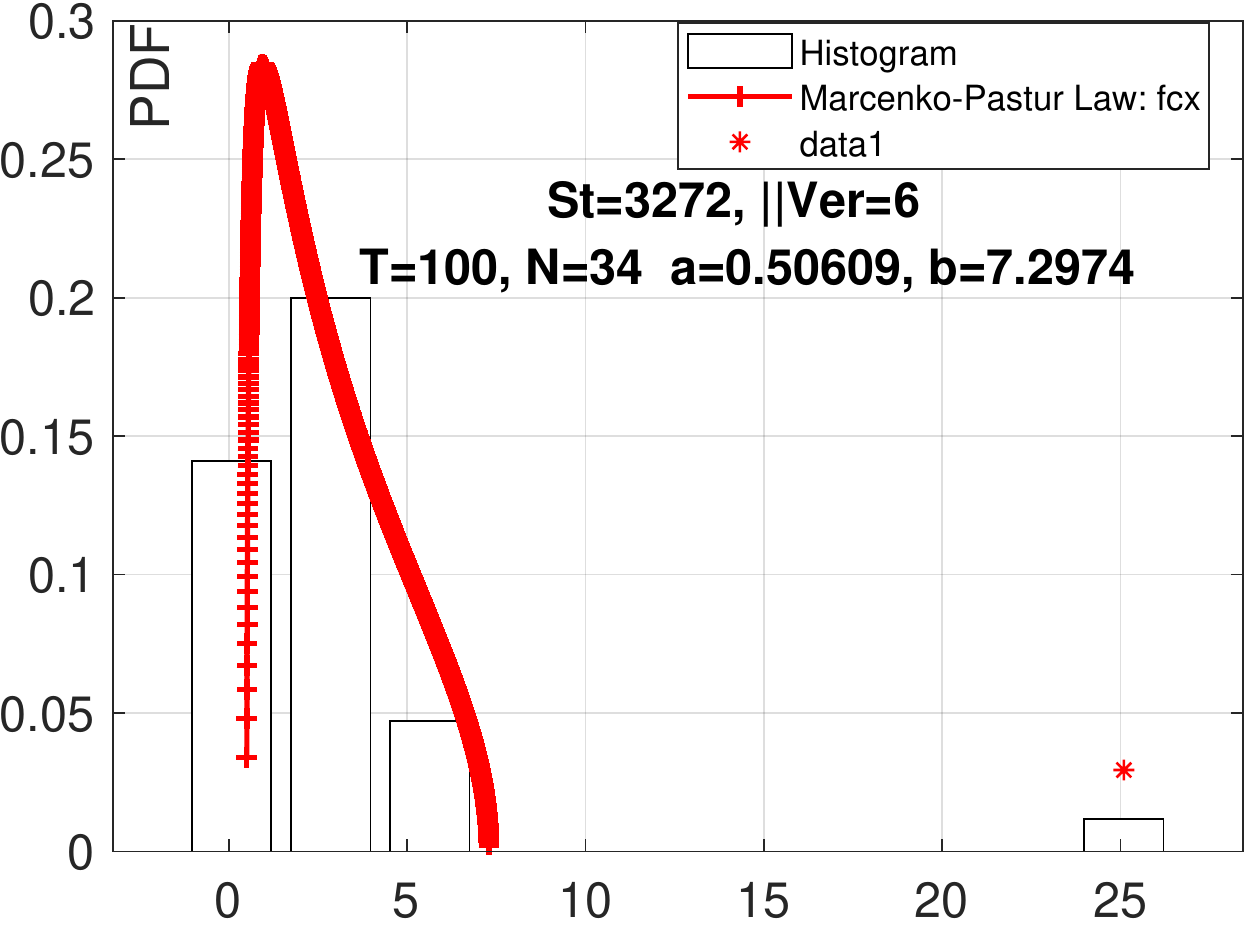}
}

\subfloat[Post-fault: Ring Law]{\label{fig:Case0T3ring}
\includegraphics[width=0.22\textwidth]{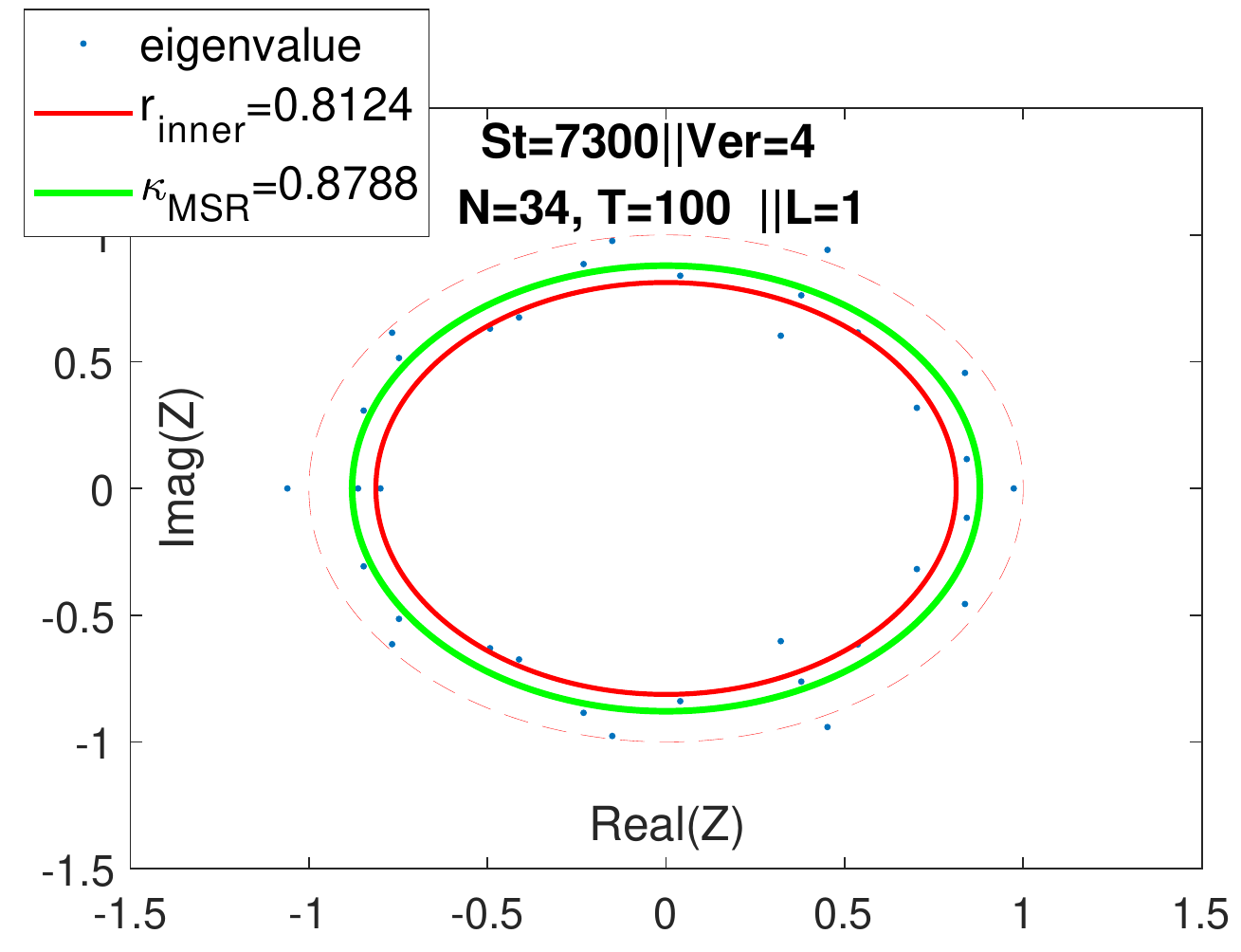}
}
\subfloat[Post-fault: M-P Law]{\label{fig:Case0T3mp}
\includegraphics[width=0.20\textwidth]{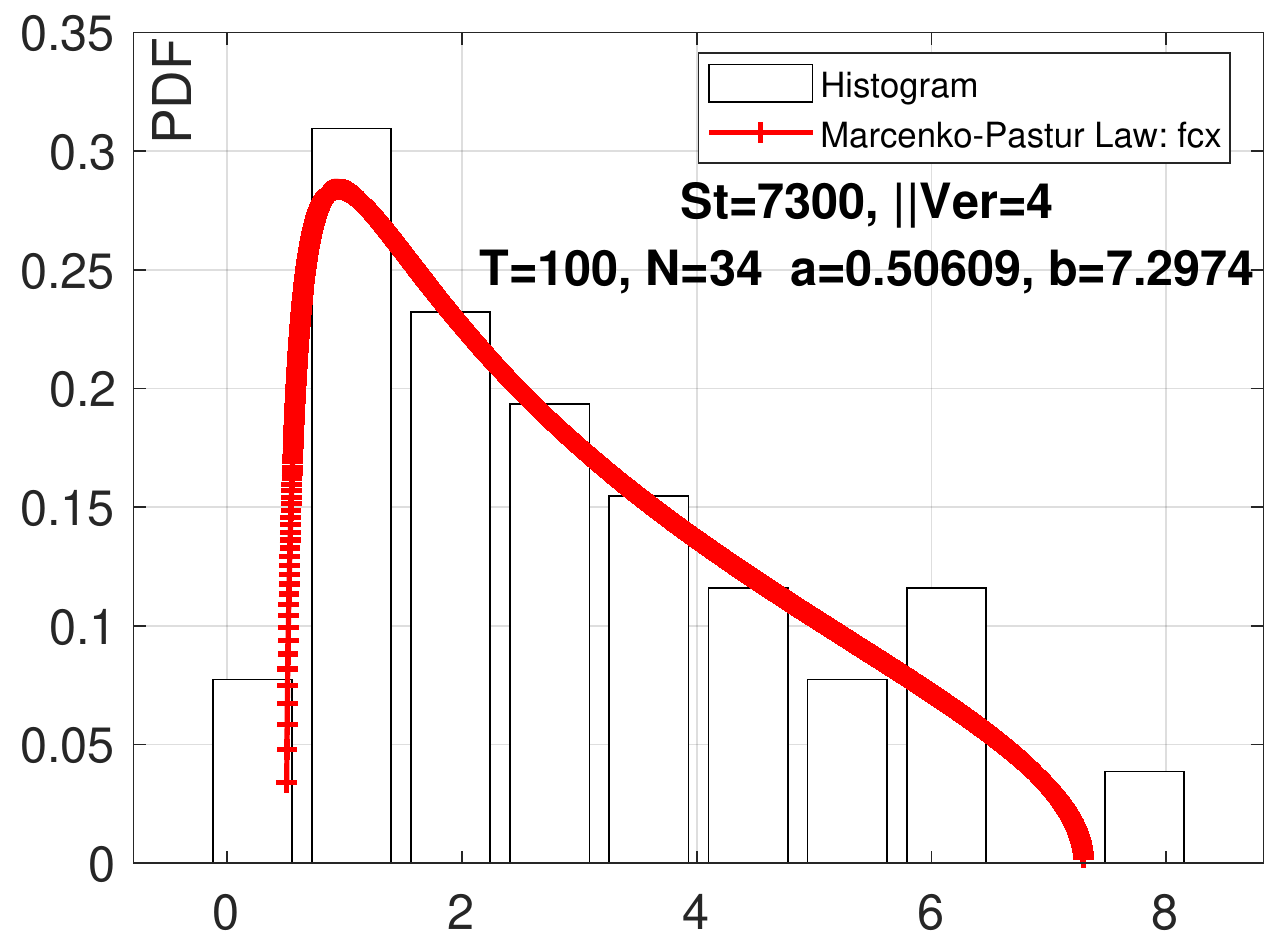}
}

\caption{Ring Law and M-P Law for the fault.}
\label{Fig:Case2GridA}
\end{figure}

\begin{figure}[H]
\centering
\includegraphics[width=0.47\textwidth]{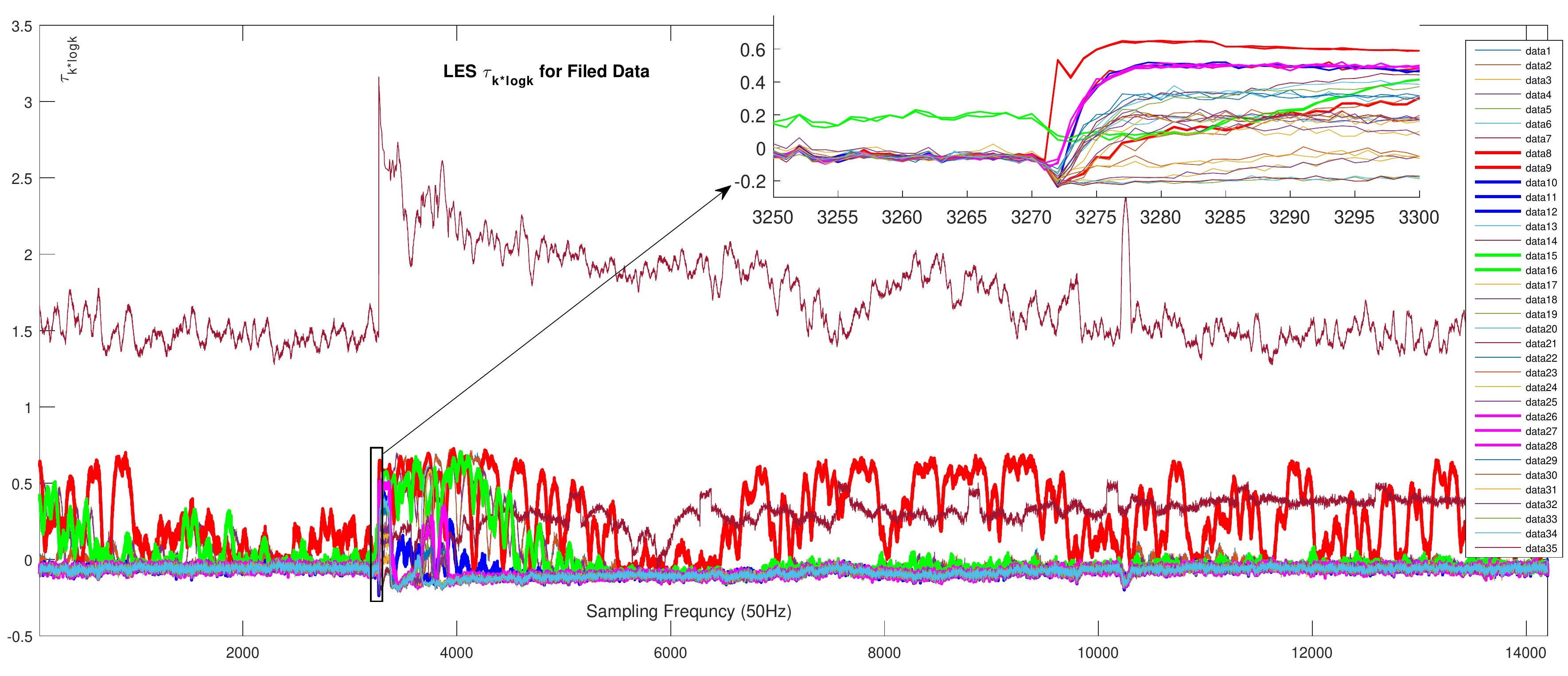}
\caption{LES $t-\tau$ curves.}
\label{fig:case2GridD}
\end{figure}

Some power grid of China are selected, with 34 PMUs collecting power flow data.
The raw data are shown as Fig.  \ref{Fig:Case2Grid}; it is quite obvious that the fault begins at sampling time $t_\text{s}\!= 3271$.
The ring distribution and M-P law pre-fault ($3101-3100$), during fault ($3173-3272$), and post-fault ($7201-7300$) are given as Fig  \ref{Fig:Case2GridA}. This implies that the real-world data do follow the Ring Law and M-P Law under normal condition, and they violate these laws when the fault is occurring. Moreover, the LES $t-\tau$ curves of basic matrix $\mathbf B$ and concatenated matrix $\mathbf C_i$ are obtain as Fig. \ref{fig:case2GridD}. It shows that Node 8, 9, 26, 27, 28, 10, 11, 12 are most relevant to this fault; while Node $1-7$ are not so sensitive.


\section{Conclusion}
\label{section: concl}
\normalsize{}
This paper has made significant progress on the basis of our previous work in the context of big data analytics for future grids. Randomness and uncertainty are at the heart of this data modeling and analysis. 
Our approach exploits the massive spatial-temporal datasets of power systems. Random matrix theory (RMT) appears very natural for the problem at hands; in a random matrix of  ${\mathbb{C}^{N \times T}},$ we use $N$ nodes to represent the spatial nodes and $T$ data samples to represent the temporal samples. When the number of nodes $N$ is large, very unique mathematical phenomenon occurs, such as concentration of measure~\cite{qiu2015smart}. Phase transition as a function of data size $N$ is a result of this deep mathematical phenomenon. This is the very reason why the proposed algorithms are so powerful in practice.

Explicitly expressed in forms of linear eigenvalue statistics (LESs)~\cite{Qiu2016BigDataLRM}, the proposed RMT-based algorithms have numerous unique advantages. They are especially suitable for complex systems. In the form of a large random matrix, they handle massive data  that are in high dimensions and within a wide time span all at once. The trick is to treat these data \textit{as a whole} at the disposal of RMT. In this way, highly reliable decisions are still attainable with some imperfect data, e.g., the asynchronous data. Moreover, with the statistical processing such as test function setting, the proposed data-driven approach has the potential to balance the perspectives of the speed, the sensitivity, and the reliability in practice.

The stability evaluation and behavior analysis are two big topics along this direction. Besides, the statistical indicators are good starting points for artificial intelligence and machine learning. For example, we can extract the linear eigenvalue statistics as features; those extracted features are used for further data processing in the pipeline using algorithms such as random forest, decision trees, and support vector machine. Our whole framework starts with the use of sample covariance matrix to replace the true covariance matrix. It is well known that this replacement is far from optimal. The almost optimal estimation of large covariance matrices using tools from  RMT~\cite{bun2016cleaning} can be used, instead.

\appendices
\section{}

\section{}
\begin{table}[H]
\caption {Series of Events}
\label{Tab: Event Series}
\centering

\begin{minipage}[htbp]{0.48\textwidth}
\centering

\begin{tabularx}{\textwidth} { >{\text}l !{\color{black}\vrule width1pt}    >{$}l<{$}|  >{$}l<{$}| >{$}l<{$}|   >{$}l<{$} }  
\toprule[1.5pt]
\hline
\textbf {Stage} & \textbf{E}1 & \textbf{E}2 & \textbf{E}3 & \textbf{E}4\\
\toprule[1pt]
Time (s) & $1--500$ & $501--900$ & $901--1300$ & $1301--2500$ \\
\hline
\VPbus{52} (MW) & 0 & 30  & 120 &  t/4-205\\
\toprule[1pt]
\end{tabularx}
\raggedright
\small {$P_{52}$ is the power demand of node 52.
}
\end{minipage}
\end{table}

The power demand of other nodes are assigned as
\FuncC{eq:gridfluctuation}{
\Equs {\Index {\Tdata y}{\Text{load\_}nt}}  {
\Index  y{\Text{load\_}nt}\Mul{}{}(1+\Muls{\Vgam{\Text {Mul}}}{r_1})
} +   \Muls{\Vgam{\Text {Acc}}}{r_2},
} 		
where $r_1$ and $r_2$ are the element of standard Gaussian random matrix;  \Vgam{\Text {Acc}}=0.1, \Vgam{\Text {Mul}}=0.001.


\small{}
\bibliographystyle{IEEEtran}
\bibliography{helx2}

\normalsize{}
\end{document}